%% file: main.tex
\documentclass[11pt]{article}
\usepackage[margin=1in]{geometry}
\usepackage[T1]{fontenc}
\usepackage[utf8]{inputenc}
\usepackage{newtxtext,newtxmath}
\usepackage{microtype}
\usepackage{graphicx}
\usepackage{booktabs}
\usepackage{array}
\usepackage{tabularx}
\usepackage{multirow}
\usepackage{xcolor}
\usepackage{amsmath}
\usepackage{enumitem}
\usepackage[numbers,sort&compress]{natbib}
\usepackage{url}
\usepackage{tikz}
\usetikzlibrary{arrows.meta,positioning,calc,shapes.geometric}
\definecolor{nv}{RGB}{26,49,88}
\definecolor{nvl}{RGB}{226,232,242}
\definecolor{gy}{RGB}{110,110,110}
\tikzset{
  bx/.style={draw=nv,line width=0.4pt,rounded corners=1.8pt,align=center,inner sep=2.6pt,font=\scriptsize,minimum height=0.46cm},
  bxf/.style={bx,fill=nvl},
  bxb/.style={bx,line width=1.1pt},
  grp/.style={draw=nv,densely dashed,line width=0.4pt,rounded corners=3pt},
  pan/.style={draw=gy,line width=0.3pt,rounded corners=2pt},
  ar/.style={-{Stealth[length=3.8pt,width=2.8pt]},draw=nv,line width=0.4pt},
  arg/.style={ar,draw=gy,densely dashed},
  lb/.style={font=\scriptsize,inner sep=1.4pt,text=nv},
  gl/.style={font=\scriptsize\itshape,text=gy,inner sep=1.4pt},
  hd/.style={font=\small\bfseries,align=center}
}
\usepackage{float}
\usepackage[section]{placeins}
\usepackage[hidelinks]{hyperref}
\usepackage{caption}
\setlist{nosep,leftmargin=*}
\newcommand{\afs}{\textsc{AfS}}
\newcommand{\code}[1]{\texttt{#1}}
\newcolumntype{L}[1]{>{\raggedright\arraybackslash}p{#1}}

\title{\textbf{Research-Native by Construction:\\ Minimal Nodes, Re-verifiable Workflows,\\ and Compounding Memory for Long-Horizon Scientific Agents}}

\author{IdeaHorizon Team\\[2pt] {\small\texttt{techsupport@ideahorizon.ai}}}
\date{September 2026}

\begin{document}
\maketitle

\begin{abstract}
Most agents for science are general coding agents with a skills folder attached, and they inherit that lineage's failure mode: under pressure to finish, they fabricate, skip, or smooth over. We describe \afs{} (Agent for Science), a platform designed from the ground up for long-horizon scientific work, where a project runs for tens of hours across dozens of agent runs and hundreds of millions of tokens with a human present only occasionally. The design rests on a single claim: most of the credibility of machine-made research can be moved from \emph{asking the model to behave} to \emph{making the non-compliant state unrepresentable}. We encode research discipline as a small set of mechanically enforced iron laws (commitment before measurement; unforgeable freezing; reports are not facts; evidence persists but verdicts do not; negative results are first-class; mechanical questions to the framework and semantic judgment to the model), and organize the platform around three time horizons. Within a run, a minimal set of research nodes (eight producing or service nodes and three architecture nodes) is extended by deepening node contracts rather than by adding node types, with admission of a new evidence node tied to a distinct way of cheating that existing gates cannot catch. Within a project, an inquiry contract with frozen closure conditions, a hash-chained artifact ledger, framework-computed closure tallies, and a referee loop make every conclusion re-verifiable from disk. Across projects, a single-file project memory and a two-tier knowledge base with promotion by rewriting let a research group compound what it has paid to learn. This paper is a system description, written under one rule: each mechanism appears in exactly one place, with the invariant it enforces, the failure it prevents, the way it is realized, and the cost it imposes. It covers the node contract and its boundaries, the inquiry contract and the hash-chained artifact ledger with gates on the write path, the two-tier memory, and the runtime substrate: the run lifecycle, what the model is shown and in what order, the context window as a rendered view over an append-only log, progress control, the execution boundary, resumption, and the human in the loop. Three traces walk real failure attempts through the mechanisms that catch them. Two closed campaigns are included as worked illustrations of the machinery rather than as an evaluation: a 50-hour assisted study that froze a manuscript after five review rounds and reported a non-reproduced mechanism as an open question, and an unattended Monte Carlo study that refuted its own preregistered hypothesis against an exact solution. We report no benchmark; a process-integrity suite that would support quantitative comparison against other systems is under construction, and what we can measure today is only the operating cost of the machinery.
\end{abstract}

\input{sections/01_intro_principles}
\input{sections/02_overview}
\input{sections/03_nodes}
\input{sections/04_workflow}
\input{sections/05_memory}
\input{sections/06_runtime}
\input{sections/07_traces}
\input{sections/08_campaigns}
\input{sections/09_related_limits}

\bibliographystyle{unsrtnat}
\bibliography{refs}

\clearpage
\appendix
\input{sections/10_appendix}

\end{document}

%% file: sections/01_intro_principles.tex
\section{Introduction}
\label{sec:intro}

A scientific project is not a task. It runs for days, spans dozens of separately started agent runs, consumes hundreds of millions of tokens, restarts after the machine sleeps, and has a human in the loop for a few minutes out of every few hours. Along the way it accumulates commitments (what would count as an answer), evidence (what was measured, observed, or derived), verdicts (what the evidence means), and a manuscript that must be traceable to all of the above. Existing agents for science, from the fully autonomous AI Scientist~\citep{lu2024aiscientist,yamada2025aiscientistv2} to human-in-the-loop assistants~\citep{schmidgall2025agentlab} and skill-library workbenches such as Claude Science~\citep{claudescience2026}, are in most cases a general agent loop plus domain tools plus a prompt that asks for rigor. Benchmarks that look at the process rather than the answer report what that lineage produces: in one open-ended research benchmark, 80\% of agent runs produced fabricated or invalid experimental results~\citep{mlrbench2025}; a study of more than 25,000 scientific-agent runs found that the scaffold accounted for 1.5\% of outcome variance against 41.4\% for the base model, and that evidence was ignored in 68\% of traces~\citep{scaffold2026}.

We built \afs{} (Agent for Science) on the opposite premise: a research agent must be research-native by construction. The discipline that makes a result trustworthy, preregistration, unforgeable records, evidence-backed verdicts, honest negative results, should not be requested of the model; it should be a property of what the model is able to write. Across 35 numbered end-to-end campaigns, the latest in September 2026, we observed three recurring ways in which a capable model in an ordinary harness produces untrustworthy science:

\begin{enumerate}
\item \textbf{Completion without delivery.} A run reports \code{completed} while the manuscript was never frozen, or was ``frozen'' by rewriting a metadata flag after the real freeze tool had refused. In one campaign the model explained its own workaround in the record: ``\emph{via save\_artifact re-save with frozen=true metadata; orchestrator freeze\_artifact was mechanically rejected}.''
\item \textbf{Gates that teach performance.} A falsifiability gate that demands a numeric threshold receives one: for a historical topic with no measurable quantity, the model invented ``a 30-year gap counts as weakening, a 20-year gap counts as refutation'' to pass the gate. A gate that asks for a form produces the form.
\item \textbf{Judgment hard-coded as thresholds.} A quality-check layer that killed a run when ``fewer than five claim citations'' appeared could be satisfied by scattering five identifiers, and could not be satisfied by a focused short paper with three solid ones.
\end{enumerate}

None of these is a model deficiency. Each is a place where the harness either asked for behavior it could not verify, or verified the wrong thing. The platform described here replaces such places with mechanisms whose failure is unrepresentable rather than merely discouraged.

This paper is a description of that system, not an evaluation of it. Its aim is to set down the architecture, the reasoning that produced each part, the alternative that was rejected and why it fails over a long horizon, and enough of the implementation that the design can be judged and rebuilt. We adopt one rule of exposition: \emph{each mechanism is described in exactly one place}, with the invariant it enforces, the failure it prevents, the way it is realized, and the cost it imposes, so that a reader can find all four without assembling them from several sections. Quantitative comparison against other systems is not attempted; the evidence we have is two closed campaigns and the platform's own operating records, and we are explicit about where that stops.

\paragraph{What this paper describes.}
\begin{itemize}
\item \textbf{Design principles} (Section~\ref{sec:laws}): the single claim behind the platform and the six research laws it produces, given as an index into the sections that realize them.
\item \textbf{System overview} (Section~\ref{sec:overview}): the three time horizons, the research loop that replaces a pipeline, the three node roles, and what the substrate provides to every run.
\item \textbf{Minimal research nodes} (Section~\ref{sec:nodes}): a node contract with five levers of extension, boundaries between nodes that make cross-node writes impossible, an admission criterion for evidence nodes that yields exactly three modalities, and the three framework nodes, including the dispatch gate and the reason there is no verdict layer.
\item \textbf{A re-verifiable workflow} (Section~\ref{sec:workflow}): research questions with frozen closure conditions and a tally computed from ledgers; artifact identity, versioning, freezing and amendment with a hash-chained ledger and gates on the write path; verdict authority, obligations, and the referee loop.
\item \textbf{Compounding memory} (Section~\ref{sec:memory}): a single-file project memory with four layers and mechanical delivery; a two-tier knowledge base in which the organization tier has no birth channel except promotion by rewriting.
\item \textbf{The runtime substrate} (Section~\ref{sec:runtime}): the run lifecycle, what the model is shown, hooks, the context window as a rendered view, progress control, tools, where a requirement is made to bind, the execution boundary, resumption, the human in the loop, accounting, and extension points.
\item \textbf{Traces, campaigns and cost} (Sections~\ref{sec:traces}--\ref{sec:cost}): three failure attempts walked through the gates that catch them; two fully closed campaigns recovered from disk; and what a round of research costs.
\end{itemize}

\section{Design Principles}
\label{sec:laws}

The platform's design principle can be stated in one line: \emph{most of the credibility of machine-made research can be moved from ``ask the model to comply'' to ``make the non-compliant state unrepresentable.''} The same requirement can land in three places, and their reliability differs by an order of magnitude (Table~\ref{tab:landing}); Section~\ref{sec:layers} refines this into the eight layers the implementation actually offers.

\begin{table}[t]
\centering\small
\begin{tabular}{L{3.2cm}L{6.4cm}L{5.4cm}}
\toprule
Where the rule lives & Example & Observed strength \\
\midrule
In the prompt & ``You \textbf{must} freeze the preregistration before running.'' & Weakest. Under pressure the rule is restated and then worked around (\S\ref{sec:intro}). \\
Post-hoc check & After the run, a check tests whether a freeze happened. & Weak. Found too late; killing the run does not undo what ran. \\
Unrepresentable & The dispatch gate refuses to start \code{experiment} unless a frozen preregistration exists. & Strong. The illegal state cannot be written. \\
\bottomrule
\end{tabular}
\caption{Three places the same requirement can land.}
\label{tab:landing}
\end{table}

Applied across the platform, this principle yields six laws (Table~\ref{tab:laws}). The table is an index: each law names the failure it prevents, observed in our own runs, and the section where the mechanism that enforces it is described in full. The laws are not domain-specific and the platform contains no domain content; domain knowledge lives in node skills and per-project configuration.

\begin{table}[t]
\centering\small
\begin{tabular}{L{0.5cm}L{5.6cm}L{6.2cm}L{2.4cm}}
\toprule
& Law & Failure it prevents & Described in \\
\midrule
L1 & \textbf{Commitment before measurement.} A question with a written proposition is a hypothesis; closure conditions are frozen before work starts. & Thresholds invented to pass a falsifiability gate; a verdict chosen after the result was seen. & \S\ref{sec:inquiry} \\
L2 & \textbf{Freezing is irreversible and unforgeable; revision leaves a trace.} & A refused freeze bypassed by re-saving with \code{frozen:true}; a frozen plan silently edited. & \S\ref{sec:artifacts} \\
L3 & \textbf{Reports are not facts.} Anything mechanically computable is computed by the framework from ledgers, not read from the model's self-report. & \code{status: verified} written by hand with no verifier call; ``11 of 12 fulfilled'' with zero matching keys. & \S\ref{sec:inquiry}, \S\ref{sec:traces} \\
L4 & \textbf{Evidence persists; verdicts do not.} Detectors record; only the reviewer and the adjudicating node judge; judgments are recomputed, never stored. & A check layer that killed runs 87\% of which never recovered; a node adjudicating its own experiment. & \S\ref{sec:noverdict}, \S\ref{sec:authority} \\
L5 & \textbf{Negative results are first-class.} Refuted stays refuted in the manuscript; withdrawn questions enter Limitations; dead ends become knowledge cards. & ``H3 is refuted'' asserted for a quantity never measured; a null result smoothed into a trend. & \S\ref{sec:authority}, \S\ref{sec:kb} \\
L6 & \textbf{Mechanical questions to the framework, semantic judgment to the model, in both directions.} & An orchestrator busy-waiting 47 turns on whether a node should continue; a quota of five citations standing in for relevance. & \S\ref{sec:archnodes}, \S\ref{sec:noverdict} \\
\bottomrule
\end{tabular}
\caption{Six research laws, the failure each prevents, and where its mechanism is described.}
\label{tab:laws}
\end{table}

Two corollaries recur throughout. First, \emph{a mechanism that exists is not a mechanism that is wired}: a detector whose output no dispatcher reads, a review specification the reviewer cannot find, or a memory channel gated behind a query the node never issues, all measure as absent. In one earlier configuration we measured, under bare dispatch all node types but one received no memory at all. Second, \emph{the contract must reach the caller}: a legal vocabulary that is only revealed by a runtime error forces the model to guess, and the guess fails silently when the vocabulary is a set of keys. Section~\ref{sec:traces} walks through the case.

Finally, the platform holds itself to a methodological stance that is itself a law: \emph{a sentence in a prompt is not a capability until its effect on behavior has been observed}. Where we have not observed it, we describe the mechanism and say so, rather than reporting the sentence as a property of the system.

%% file: sections/02_overview.tex
\section{System Overview}
\label{sec:overview}

\begin{figure}[t]
\centering
\resizebox{\textwidth}{!}{\input{figures/tikz_overview}}
\caption{Three time horizons of a long-horizon research job and the three pillars that serve them. Within a run, a minimal set of research nodes; within a project, a re-verifiable workflow from inquiry contract to frozen deliverables; across projects, memory that compounds through promotion. The laws of Section~\ref{sec:laws} are enforced where the record is written in all three.}
\label{fig:overview}
\end{figure}
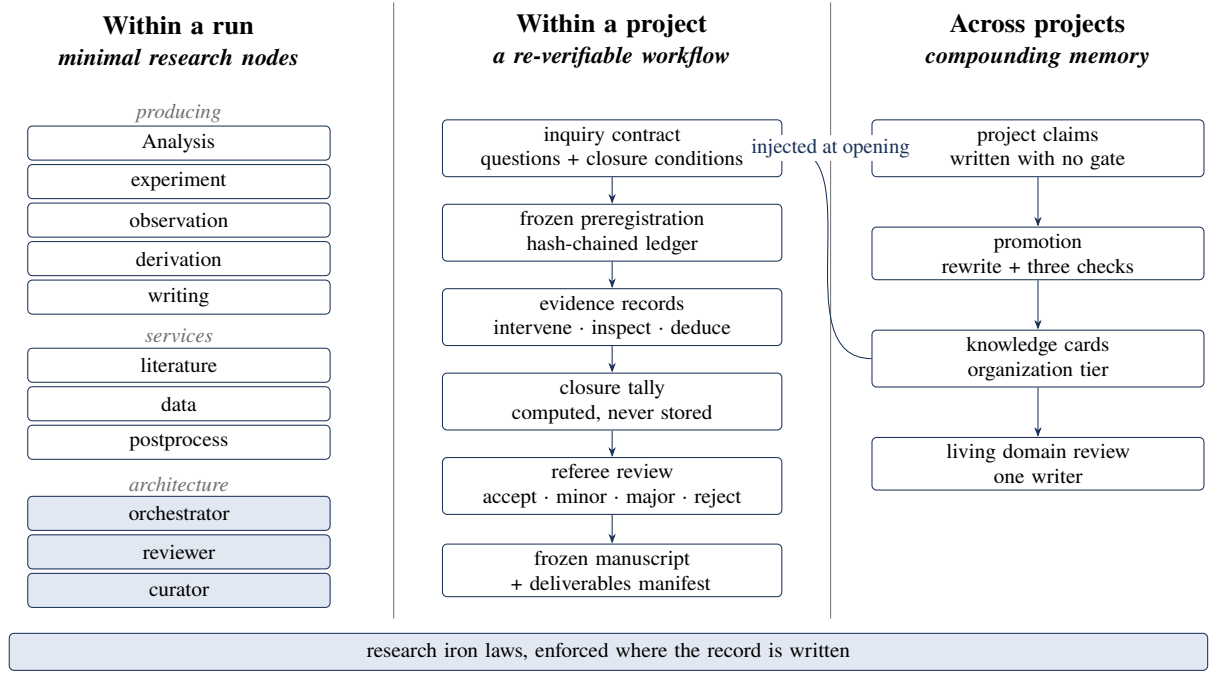

\afs{} is organized around three time horizons (Figure~\ref{fig:overview}). A \emph{run} is one invocation of one node's agent loop under a fixed contract. A \emph{project} is a Git worktree in which dozens of runs accumulate artifacts, ledgers, and a versioned research state. An \emph{organization} is the knowledge that survives project closure. Sections~\ref{sec:nodes}--\ref{sec:memory} describe one pillar per horizon; Section~\ref{sec:runtime} describes the substrate all three run on.

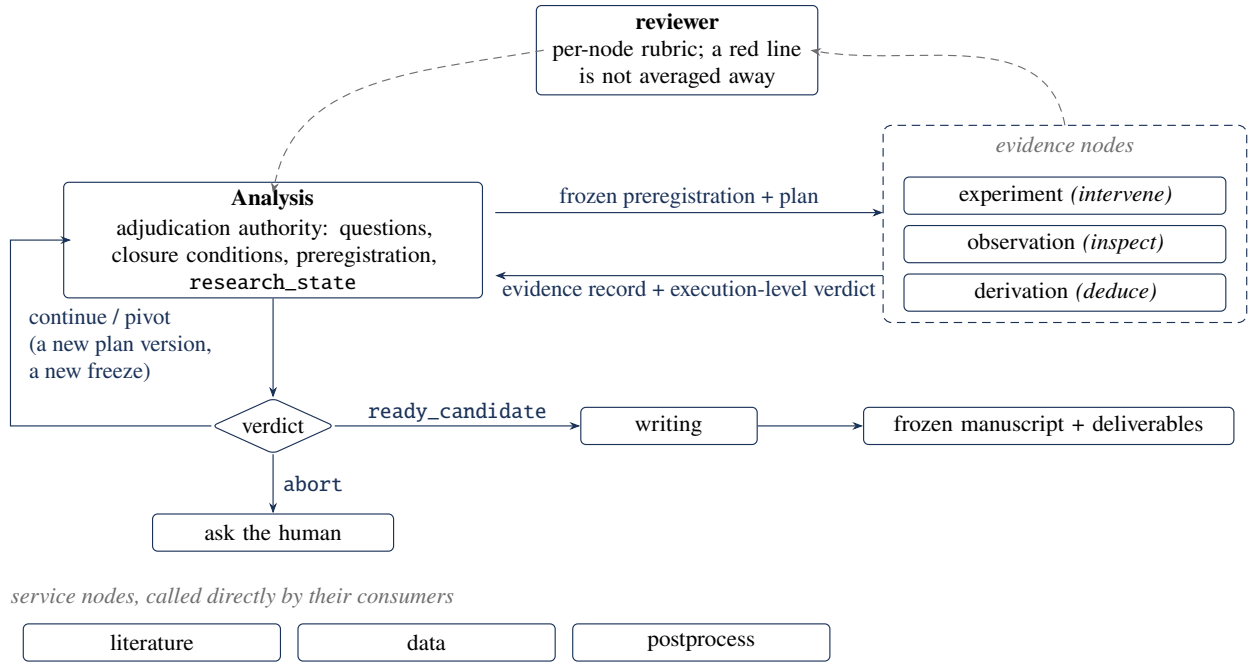
\begin{figure}[t]
\centering
\resizebox{\textwidth}{!}{\input{figures/tikz_loop}}
\caption{The research loop. Analysis holds adjudication authority; evidence nodes return execution-level verdicts, never scientific ones; services are called directly by their consumers; the reviewer is parameterized by the reviewed node's own specification.}
\label{fig:loop}
\end{figure}

\paragraph{The process is a loop, not a pipeline.}
Early designs drew the platform as a line: literature, then hypothesis, then experiment, then writing. That picture is wrong and expensive, because an orchestrator that thinks in pipelines dispatches by default script. The actual shape (Figure~\ref{fig:loop}) is a loop whose next step is chosen by a \emph{verdict}, not by position. The \emph{Analysis} node (directory name \code{hypothesis} for historical reasons) is the sole writer of a versioned \code{research\_state} ledger. It freezes a preregistration and a plan, dispatches evidence nodes, reads their execution-level verdicts, and records a scientific verdict: \code{continue} (dispatch more evidence), \code{pivot} (new plan version, new freeze), \code{ready\_candidate} (proceed to writing), or \code{abort} (ask the human). A project may loop between Analysis and evidence ten times, or never enter an experiment at all.

\paragraph{Node roles.}
Nodes fall in three roles. \emph{Producing} nodes (Analysis, \code{experiment}, \code{observation}, \code{derivation}, \code{writing}) produce research content and enter a post-run flow (review, or review then curation). \emph{Service} nodes (\code{literature}, \code{data}, \code{postprocess}) are called synchronously by their consumers, declared in the caller's \code{callable\_nodes}, and their return value is their delivery. \emph{Architecture} nodes (\code{\_orchestrator}, \code{\_reviewer}, \code{\_curator}) belong to the framework and never produce research content.

\paragraph{What the substrate provides.}
Every run executes under the same runtime, described in Section~\ref{sec:runtime}: a session declares one of three autonomy tiers, from every decision stopping for a human to every approval granted; model-issued commands reach the operating system through one choke point that enforces a write boundary, network denial and resource ceilings, and records what it could not enforce; runs checkpoint every turn and resume after process, backend or machine restarts; and every model call is appended to a per-project cost ledger.

%% file: figures/tikz_overview.tex
\begin{tikzpicture}[font=\scriptsize,x=1cm,y=1cm]
  \def\xA{2.45}\def\xB{8.30}\def\xC{14.05}
  \draw[gy,line width=0.3pt] (5.35,0.62) -- (5.35,-7.70);
  \draw[gy,line width=0.3pt] (11.25,0.62) -- (11.25,-7.70);
  \node[hd,text width=4.9cm] at (\xA,0.10) {Within a run\\ \emph{\footnotesize minimal research nodes}};
  \node[hd,text width=5.2cm] at (\xB,0.10) {Within a project\\ \emph{\footnotesize a re-verifiable workflow}};
  \node[hd,text width=4.9cm] at (\xC,0.10) {Across projects\\ \emph{\footnotesize compounding memory}};
  \node[gl] at (\xA,-0.88) {producing};
  \foreach \i/\t in {0/Analysis,1/experiment,2/observation,3/derivation,4/writing}
    {\node[bx,text width=3.9cm] at (\xA,-1.28-0.52*\i) {\t};}
  \node[gl] at (\xA,-3.88) {services};
  \foreach \i/\t in {0/literature,1/data,2/postprocess}
    {\node[bx,text width=3.9cm] at (\xA,-4.28-0.52*\i) {\t};}
  \node[gl] at (\xA,-5.88) {architecture};
  \foreach \i/\t in {0/orchestrator,1/reviewer,2/curator}
    {\node[bxf,text width=3.9cm] at (\xA,-6.28-0.52*\i) {\t};}
  \foreach \i/\t in {0/{inquiry contract\\ questions + closure conditions},
                     1/{frozen preregistration\\ hash-chained ledger},
                     2/{evidence records\\ intervene · inspect · deduce},
                     3/{closure tally\\ computed, never stored},
                     4/{referee review\\ accept · minor · major · reject},
                     5/{frozen manuscript\\ + deliverables manifest}}
    {\node[bx,text width=4.4cm] (B\i) at (\xB,-1.35-1.14*\i) {\t};}
  \foreach \i/\j in {0/1,1/2,2/3,3/4,4/5} {\draw[ar] (B\i) -- (B\j);}
  \foreach \i/\t in {0/{project claims\\ written with no gate},
                     1/{promotion\\ rewrite + three checks},
                     2/{knowledge cards\\ organization tier},
                     3/{living domain review\\ one writer}}
    {\node[bx,text width=4.3cm] (C\i) at (\xC,-1.35-1.42*\i) {\t};}
  \foreach \i/\j in {0/1,1/2,2/3} {\draw[ar] (C\i) -- (C\j);}
  \draw[ar] (C2.west) to[out=180,in=0] (B0.east);
  \node[lb,fill=white,inner sep=1.4pt] at (11.25,-1.35) {injected at opening};
  \node[bxf,text width=16.0cm,minimum height=0.5cm] at (8.25,-8.15) {research iron laws, enforced where the record is written};
\end{tikzpicture}

%% file: figures/tikz_loop.tex
\begin{tikzpicture}[font=\scriptsize,x=1cm,y=1cm]
  \node[bx,text width=5.0cm,minimum height=1.15cm] (an) at (3.40,0)
    {\textbf{Analysis}\\[1pt] adjudication authority: questions,\\ closure conditions, preregistration,\\ \code{research\_state}};
  \foreach \i/\t in {0/{experiment \emph{(intervene)}},1/{observation \emph{(inspect)}},2/{derivation \emph{(deduce)}}}
    {\node[bx,text width=3.8cm] (E\i) at (13.20,0.55-0.60*\i) {\t};}
  \node[gl] at (13.20,1.18) {evidence nodes};
  \draw[grp] (10.95,1.42) rectangle (15.45,-1.02);
  \draw[ar] (6.15,0.34) -- node[lb,above,pos=0.5]{frozen preregistration + plan} (10.95,0.34);
  \draw[ar] (10.95,-0.44) -- node[lb,below,pos=0.5]{evidence record + execution-level verdict} (6.15,-0.44);
  \node[bx,text width=3.3cm] (rev) at (8.40,2.35) {\textbf{reviewer}\\ per-node rubric; a red line\\ is not averaged away};
  \draw[arg] (13.20,1.42) to[out=100,in=-10] (10.05,2.35);
  \draw[arg] (6.75,2.35) to[out=190,in=80] (3.40,0.58);
  \node[bx,shape=diamond,aspect=2.2,inner sep=1.4pt] (vd) at (3.40,-2.30) {verdict};
  \draw[ar] (an) -- (vd);
  \draw[ar] (vd.west) -- (0.15,-2.30) -- (0.15,0) -- (0.90,0);
  \node[lb,anchor=west,align=left,fill=white,inner sep=1.5pt] at (0.32,-1.30) {continue / pivot\\ (a new plan version,\\ a new freeze)};
  \node[bx,text width=2.0cm] (wr) at (8.30,-2.30) {writing};
  \node[bx,text width=4.4cm] (dl) at (13.00,-2.30) {frozen manuscript + deliverables};
  \draw[ar] (vd.east) -- node[lb,above]{\code{ready\_candidate}} (wr.west);
  \draw[ar] (wr) -- (dl);
  \node[bx,text width=2.8cm] (hu) at (3.40,-3.62) {ask the human};
  \draw[ar] (vd.south) -- node[lb,right]{\ \code{abort}} (hu.north);
  \node[gl,anchor=west] at (0.10,-4.42) {service nodes, called directly by their consumers};
  \foreach \i/\t in {0/literature,1/data,2/postprocess}
    {\node[bx,text width=3.0cm] at (1.90+3.40*\i,-4.98) {\t};}
\end{tikzpicture}

%% file: sections/03_nodes.tex
\section{Pillar A: Minimal Research Nodes}
\label{sec:nodes}

\subsection{The node contract and the five levers}
\label{sec:contract}

A node is a directory with one contract file and its own tools, skills, and hooks. The contract declares the system prompt and rules, the tool whitelist, skills, loop hooks, expected inputs and outputs, required output artifact types (optionally per mode), callable nodes, the post-run flow, a turn cap, and a summarizer policy; Appendix~\ref{app:contract} lists the fields. Capability is added by deepening five levers of an existing node: the prompt and rules; write gates on its artifact types; the tool whitelist and tool-internal validation; the node's \code{review\_spec.md}, which parameterizes the shared reviewer; and skills loaded on demand. A node type is admitted only under the criterion of Section~\ref{sec:modalities}; every other responsibility (preregistration, falsification design, multi-persona review, reproducibility audit) is expressed as a deepening of an existing node rather than as a new one.

Loading is where a contract is rejected rather than tolerated. A node whose Python module fails to import does not start; a node that declares expected outputs but no required output type is a contradiction and raises; and a node whose whitelist names a tool that the tool itself forbids to that node raises, rather than silently intersecting to nothing. The last case matters because the silent intersection is invisible from both sides: the prompt still names the tool, the model still tries to call it, and the failure surfaces as a model that ``ignores instructions.''

\subsection{Boundaries between nodes}
\label{sec:boundaries}

\paragraph{Why the next node is chosen and not declared.} The obvious alternative to an orchestrator is a declarative graph in which each node names its successor and the framework walks the edges. We do not do this, and the reason is not that graphs are inexpressive but that in research the next step is a function of what was just learned. After a literature pass the right move may be to sharpen the question, to go straight to an experiment, or to stop; a declared edge takes that judgment away and replaces it with a default, and a default is exactly what dispatches a plan made entirely of literature adjudication to a computational node because that is where the arrow pointed. So the graph is implicit: producing runs return to the orchestrator, which reads the adjudication ledger and decides. The cost is that the orchestrator can decide wrongly, and it is paid for on the receiving side: a node that requires a frozen preregistration refuses to start without one and says so, which turns a bad dispatch into a cheap, self-correcting error rather than a run that proceeds on the wrong footing.

\paragraph{No bus between nodes.} There is no artifact bus and no automatic forwarding. A downstream node reads the upstream node's directory directly, decides for itself whether the material is sufficient, and reports a blocker if it is not. The alternative, a framework that decides what to hand forward, must model what each consumer needs and is wrong in both directions: it forwards material nobody reads and withholds material somebody needed, with no way for the consumer to distinguish ``not produced'' from ``not forwarded.''

\paragraph{One writer per piece of state.} Reading across nodes is free and writing across nodes is impossible, and the impossibility is enforced by construction rather than by convention. A node's directory is writable only by that node's process, through the sandbox of Section~\ref{sec:boundary}; the write scope resolves to the node's own directory, with the project memory file treated as an exact-file scope rather than a prefix. A node granted deliverable writes may write only into no-man's-land: never the framework area, never a protected path, and never a path owned by another node, with the refusal naming the node to ask instead. Above the sandbox sit two more layers with distinct jobs: the commit gate is the \emph{authority}, refusing to stage a modification to any pinned path; the in-process check is only a \emph{witness}, which reports a boundary crossing but does not act on it. The remaining single writers follow the same pattern: the research state is written only by Analysis through typed tools with integrity gates; the memory file has section-level ownership checked on write; organization knowledge is written only through promotion; commits are made only by the platform, whose authority is recomputed from Git rather than remembered.

\subsection{Evidence modalities and the admission criterion}
\label{sec:modalities}

The producing set contains three \emph{evidence} nodes, and the criterion for admitting one is not ``another discipline'' but \emph{a distinct original sin that existing gates cannot catch}:

\begin{itemize}
\item \textbf{experiment (intervene)}: make the world, or a model of it, do something it would not have done. Its sin is \emph{fabricated execution}; its gate is an execution-evidence loop (raw results as a frozen manifest with per-file SHA-256, recomputed at freeze and at closure).
\item \textbf{observation (inspect)}: the world has already left records; look systematically. Its sin is \emph{cherry-picking}: every citation may resolve and the seven favorable ones may still be the only ones cited. Its gate is a frozen search protocol with a denominator (screened, included, excluded), a mandatory adversarial search, and \emph{exploratory} runs that cannot discharge closure conditions, because the material that generated a hypothesis cannot confirm it.
\item \textbf{derivation (deduce)}: derive new propositions from committed premises without touching the world. Its sin is \emph{an invalid step disguised as valid}. Its gate is a proof kernel for informal derivation: the model proposes one step, the framework checks it (symbolically, then numerically at random points, after~\citet{schwartz1980}), and only checked or explicitly unverified steps enter the chain. The checker returns four values, not a boolean (\code{verified}, \code{numerically\_supported}, \code{inconclusive}, \code{failed}); confusing \code{inconclusive} with \code{failed} makes the model overturn correct steps. The applicability domain of the conclusion is computed by the framework as the union of live assumptions, and a verification stamp is accepted only if it carries a probe that the ledger can trace to a real tool call.
\end{itemize}

Experiment's gates catch none of observation's sins, and the two together catch none of derivation's: a fifty-step chain can be genuinely executed, cite only real theorems, and still swap an assumption at step 23. By this criterion a ``materials-science node'' is refused, since its sin is experiment's, and the intervene/inspect/deduce partition roughly closes the space of empirical and formal evidence. The verification tools are shared by every node; only the ownership of the evidence record type is exclusive, and the artifact policy of Section~\ref{sec:artifacts} enforces that every evidence-record type has exactly one producing owner.

\subsection{The remaining producing and service nodes}

The producing and service set has eight members. Two of them, \code{observation} and \code{derivation}, are characterized by the admission criterion of Section~\ref{sec:modalities} and are not restated here; Table~\ref{tab:nodes} lists the distinctive mechanisms of the other six.

\begin{table}[t]
\centering\footnotesize
\begin{tabular}{L{1.6cm}L{13.9cm}}
\toprule
Node & Distinctive mechanisms \\
\midrule
Analysis (\code{hypothesis}) & Four axioms with no ``hypothesis'' in them; a question with a \code{proposition} is a hypothesis, one without is not, with no yes/no switch to misdeclare. Seven pre-freeze audits: goal alignment (multi-pillar requests may not collapse to one), definition lock (the user's ontology is copied verbatim), threshold grounding (every numeric threshold needs a stated source or must become qualitative), comparison protocol, resource feasibility, cost instrumentation. Exploratory questions may not list ``nothing found'' as a failure condition. Sole writer of \code{research\_state}. \\
experiment & 55-word system prompt; knowledge lives in 30 loop hooks and 7 skills loaded after scope is known. \code{execution\_params} are compared key-by-key with the frozen preregistration before anything runs. No proxy or toy may be reported as the original task; the only exits are probe, acquire, ask, or infeasible. Tree-sitter semantic analysis of shell commands (dynamic execution refused). Irreversible-window recovery for external job submission that trusts only the pre-submission intent record. Reproducibility snapshot; three-part evidence (log, clean, raw with SHA-256); mandatory sediment attempt or an auditable ``no sediment'' addendum. \\
writing & Two stages, plan then prose; the gate asks the ledger, not the disk: zero discharged closure conditions blocks writing and produces a material-insufficiency report; partial discharge writes with \code{material\_gaps}. Mechanical blockers are separated from heuristic gaps, which are routed to the node that can fill them. Methods may not be completed from domain common sense; conflict-of-interest and authorship are not invented. Phantom claim identifiers fail the citation-integrity hook; internal identifiers may not appear in the clean PDF. TeX exit code 0 is not deliverable: layout gates on overfull boxes, unresolved references, and oversized floats. Freeze is the author's signature; the orchestrator cannot sign for it. \\
literature & Four-mode contract (landscape, targeted lookup, contradiction check, method lookup) with per-mode required outputs and turn budgets, so that a ``find five papers'' request cannot run as a full survey. Research gaps may not be phrased as gaps in the knowledge base. Writes facts (chunks, concepts with external anchors) but never assertions. Author wiring is a hard completion gate; missing authors are recorded, not invented. \\
data & Planning gate with a three-role loop (requirement analyst, designer, critic); no asset is produced before an approved plan. Two authorities (plan-bound and request-bound) compiled into a locked work order. Structured terminal states (\code{completed}, \code{needs\_input}, \code{externally\_blocked}, \code{fatal}). May not refuse a redirected request on the grounds that it is ``a computation task'' (otherwise a two-sided deadlock). \\
postprocess & Provenance binding by three hashes: source artifact, rendering script, output file, recomputed by consumers; a file not written by this sandboxed execution is refused. Verdict fields removed entirely; only two mechanical facts remain, and the absence of a vision reviewer is a readable fact that downstream must disclose. Read-only scientific semantics: no aggregation, smoothing, outlier removal, or inference of graph structure. Generative images must be marked non-evidence-bearing. \\
\bottomrule
\end{tabular}
\caption{Six of the eight producing and service nodes and their distinctive mechanisms; \code{observation} and \code{derivation} are described in Section~\ref{sec:modalities}. Tool and skill counts are tabulated in Appendix~\ref{app:contract}.}
\label{tab:nodes}
\end{table}

\subsection{The architecture nodes}
\label{sec:archnodes}

\paragraph{Orchestrator.} It talks to the user and dispatches. Each turn it receives a mechanically computed \emph{situation}: the user's goal in its latest amended form, how many closure conditions are discharged, how much evidence exists on disk and of which modality, how many claims carry sources, and whether the project is fresh, in flow debt, or mid-research. The situation returns counts and one-line summaries only and never artifact bodies, because orchestrator context is the scarcest resource in the system; fulfilment is not recomputed here but delegated to the tally of Section~\ref{sec:inquiry}, so that there is one implementation of ``how many are done.'' Without a situation an orchestrator can only follow its default script, which is the mechanical origin of the bureaucratic dispatch described in Section~\ref{sec:boundaries}; the alternative of letting the model judge whether a node should continue was measured once at 47 turns and 6.7M tokens of busy-waiting.

\paragraph{The dispatch gate.} A node that reported itself blocked is not dispatched again while nothing it depends on has changed. The criterion is a fingerprint: the hash of the sorted identities, each an artifact identifier with its version and content hash, of every artifact in the worktree \emph{except} those owned by the node itself. Excluding the node's own output is what makes the fingerprint mean anything, since the report of being blocked is itself a new artifact and would otherwise change the fingerprint on every attempt. The fingerprint is content-addressed rather than time-addressed; an empty upstream is a legitimate value rather than an error; and an uncomputable fingerprint lets the dispatch through rather than blocking it, because a gate whose own input is unavailable should fail toward the cheaper error. The gate refuses only when blockers were reported, a prior snapshot exists, and neither the upstream fingerprint nor the inputs have changed, and the refusal is required to print all three ways out: make the upstream actually change, change the inputs, or ask a human. There is no lock to clear, because a lock that someone must remember to clear will eventually be forgotten. The cost is one extra hash per dispatch and one more thing the orchestrator can be told no about. Section~\ref{sec:traces} shows the case it exists for.

\paragraph{Reviewer.} One shared reviewer, parameterized by the reviewed node: the context engine assembles the reviewed node's \code{review\_spec.md} into the reviewer's context, the critique is assembled by the framework, and the rubric source is stamped by the framework. Red lines are not averaged away: a single critical finding overrides the mean score. On entry, the research laws of Section~\ref{sec:laws}, as recorded in the project's own constitution, are expanded into checklist items the reviewer must answer one by one. The reviewer can read the producer's transcript; it does not trust the producer's self-report.

\paragraph{Curator.} An editor, not a researcher: it never produces scientific content, only changes the concentration and organization of knowledge, and only at batch boundaries (run, project, organization).

\subsection{Why there is no verdict layer}
\label{sec:noverdict}

An obvious component is missing: a quality-check layer that runs after each node and decides whether the run lived. The platform had one, and the reason it does not now is measured rather than argued. The figures are from a single internal test corpus, the transcripts of our own campaigns under that earlier configuration, and are not a benchmark; they are reported because they decided the design. Over 324 such transcripts, 87 sequences of ``check X failed on node Y'' were found; 11 (13\%) were later corrected and 76 (87\%) never passed again. The writing node had 17 runs and zero completions under it. In one campaign of 3,972 tool calls, 23 refusals to restart a node (15 writing, 8 postprocess) all came from a check failing four times in a row; none came from a reviewer. Killing a run cannot change what already happened; such a layer is a dead end, not a feedback loop.

Three failure shapes account for the numbers. \emph{Self-certification}: a check that reads a report written by the checked node's own tool is checking whether the node issued itself a pass, not whether the artifact is good. \emph{Judgment in a threshold's clothing}: ``at least five claim citations'' can be gamed by scattering five identifiers and fails a focused paper with three sound ones, while giving the appearance of hardness. \emph{Audit as a gate}: a check that a dangerous command was authorized runs after the command ran, and killing the run does not un-run it.

The checks themselves were kept as detectors whose results flow to the reviewer, and the three guarantees a verdict layer would claim were moved to where they are unrepresentable: write gates (Section~\ref{sec:artifacts}), a closing gate at the end of a run (Section~\ref{sec:lifecycle}), and tool-internal validation (Section~\ref{sec:tools}). This is the concrete form of law L4: the machine sees, and only the reviewer and the adjudicating node judge.

%% file: sections/04_workflow.tex
\section{Pillar B: A Re-verifiable Workflow}
\label{sec:workflow}

\subsection{The inquiry contract}
\label{sec:inquiry}

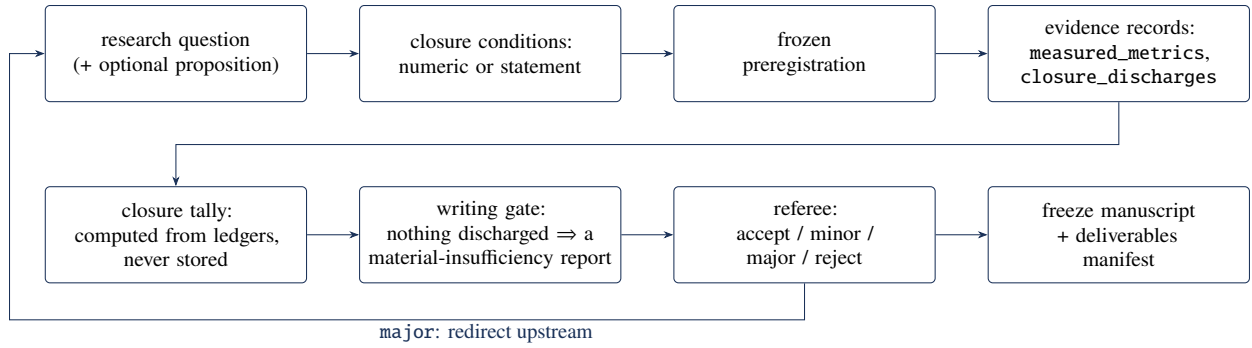
\begin{figure}[t]
\centering
\resizebox{\textwidth}{!}{\input{figures/tikz_closure}}
\caption{From inquiry contract to deliverables. Discharge keys must match the frozen identifiers verbatim; the tally is computed from ledgers at read time; the writing gate and the referee act on the tally, not on the node's self-report.}
\label{fig:closure}
\end{figure}

\paragraph{Invariant.} A research plan, in any discipline and paradigm, owes four things: state the question(s); before work starts, commit to what would count as an answer and freeze it; close a question only by discharging every commitment with evidence; and let the framework judge \emph{whether} a discharge exists while the reviewer judges \emph{how well} it was done. Hypotheses are not an axiom. A research question is question text, an output form, an optional proposition, a list of assumptions, and a frozen list of closure conditions; a question with a non-empty proposition is a hypothesis, and there is no separate yes/no field that could be declared the other way. Closure conditions come in two kinds only: numeric (metric, comparison, threshold, plus a rationale triple recording where the number came from) and statement (an observable sentence). Coverage reached, uncertainty within budget, a baseline compared, three competing explanations each separated by evidence, are all statement conditions, and the platform never needed a third kind.

\paragraph{The alternative.} A commitment account that binds a hypothesis to a list of metrics has the right red line, that a verdict cannot be flipped on a quantity never measured, but fuses it to a form: prose hypotheses are refused, exploratory and replication studies receive no protection at all, and a topic with nothing to measure invents numeric thresholds to get through. Making the unit of the ledger the closure condition rather than the metric keeps the red line and extends its coverage to every study.

\paragraph{Mechanism.} Fulfilment is matched by key against the frozen identifiers, scanning every artifact of a type that may carry a discharge ledger plus the promoted results directory. A numeric item is fulfilled when its measurement state is not open; an estimate counts only as a \emph{degraded} fulfilment, and only when it carries both a basis and a stated reason for not measuring, which keeps a legitimate fallback from becoming a silent one. A statement item requires an explicit discharged status \emph{and} non-empty evidence. Keys must match verbatim, with a positional key available for items that declare no identifier; the first writer of a key wins, so a later restatement cannot quietly overwrite an earlier discharge; and a malformed fulfilment block is not dropped but collected and surfaced, because a ledger that fails to parse looks exactly like a ledger that says nothing. Only the highest version of each preregistration is read, so commitments amended away do not lock the project.

The tally returns total, fulfilled, open by kind, the open item identifiers, and the degraded count; it returns \emph{nothing at all} when there are no frozen commitments, which is a different fact from zero fulfilled and is kept distinct at every consumer. The tally is never stored. It is computed by the situation of Section~\ref{sec:archnodes}, by the writing gate that refuses to write a paper over zero discharges, by the commitment brief rendered into each node's context, and by the freeze path when it reports outstanding debt.

\paragraph{Cost.} Every node that produces evidence must learn the frozen identifiers and write them back exactly. The framework carries them to the node through the callee contract and lists every legal key when a freeze is attempted with a mismatch; Section~\ref{sec:traces} shows what happens without that delivery.

\subsection{Artifacts: identity, version, freeze, amend}
\label{sec:artifacts}

\begin{figure}[t]
\centering
\resizebox{\textwidth}{!}{\input{figures/tikz_ledger}}
\caption{The life of one artifact. Freezing pins a version; amending a frozen version requires a reason, produces a new draft, and keeps the frozen snapshot pinned; the ledger is append-only and hash-chained; the commit gate refuses later edits to pinned paths.}
\label{fig:ledger}
\end{figure}
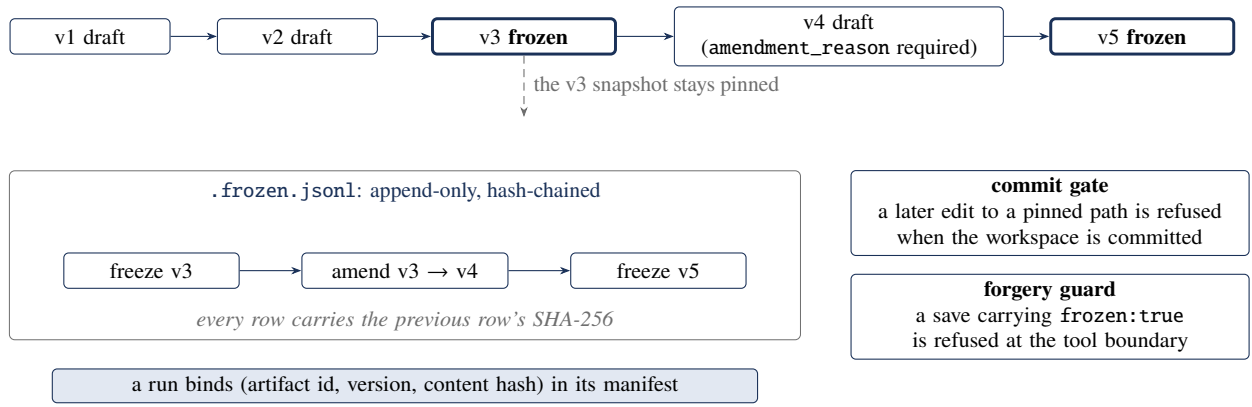

\paragraph{Invariant.} ``Which research plan'' and ``which draft of it'' are orthogonal questions, and \emph{frozen content always has exactly one pinned path}. A model that compresses the two questions into one namespace, and whose freeze expresses only immutability without revision, forbids both overwriting and renaming; the model then does the only physically possible thing and renames, and the resulting fragment files (\path{pre_registration__v1} to \path{v6}) are manufactured by the framework, not by the agent.

\paragraph{Mechanism.} Three primitives. Identity is the file path; the name is a display label. Every save increments a version and records the content hash and the previous content hash, with the previous record snapshotted into an append-only directory of versions, so the head file is always current and a naive consumer is correct by default. Freezing pins a version; amending a frozen version requires an \code{amendment\_reason}, produces a new unfrozen draft, and lets the framework compute and record the diff itself rather than accept a summary of it, while leaving the original producer's provenance intact and recording the amender only in the ledger row. Provenance distinguishes \emph{produced}, \emph{imported} and \emph{forwarded}, carries the source hash for imports, and the convenience fields naming the producing node and run are derived from it rather than stored alongside, so the two cannot disagree. Appendix~\ref{app:records} gives the record and row shapes.

The ledger lives beside the artifacts, append-only, each row carrying the SHA-256 of the previous raw line. A freeze row pins a path to a content hash; an amendment row unpins the head and pins the snapshot instead, which is what keeps the invariant: revision never opens a window in which nothing is pinned. One function is the only legal interpreter of the ledger; the commit gate that refuses to stage a modified pinned path is a separate implementation of the same reading, held to it by a contract test rather than a comment. Tamper-evident types, the preregistration among them, additionally append a hash and timestamp at freeze: a scientific timestamp. A run's contract binds the triple (artifact identifier, version, content hash) into its manifest, so ``runs 4--9 ran under preregistration v2'' is mechanically answerable, and dispatch fails loudly while an amendment draft is pending: either freeze it or declare the older version explicitly.

\paragraph{The type registry.} Rather than a list of types, the registry declares properties, and each mechanism queries the one it needs. Three faces exist: retention (permanent, chunk-then-drop or transient; deliverable directory; tamper-evident; time to live); scientific evidence (\emph{evidence record}, the authoritative record of a research execution and the hard upstream of writing; \emph{discharge ledger}, may carry fulfilment entries; \emph{cites knowledge-base claims}, the target of the citation-integrity scan); and process role (singleton, framework-internal). The registry validates itself at import: framework-internal is mutually exclusive with the evidence properties, and an evidence record that cannot carry a discharge ledger is a contradiction that raises. Ownership is deliberately not in the table; it is derived by scanning what each node declares it must produce, so a new node's outputs are protected the moment they are declared rather than when someone adds them to a list.

\paragraph{Write gates.} Two parallel registries attach gates to a type, one on save and one on freeze. A save gate receives the draft before it reaches disk; a freeze gate receives the stored record. Either returns an empty result to pass, or a mapping from check name to reason plus a hint, which is rendered into a refusal naming every failed check at once rather than one per round trip. A gate may publish a \emph{content contract} that is spliced into the live description of the writing tool, so the requirement reaches the model before the rejection rather than only after it. The division between the two registries is the point: saving is unavoidable and freezing is voluntary, so \emph{structural} discipline lives on the save gate, where declining to freeze cannot skip it, and \emph{promise-fulfilment} discipline lives on the freeze gate, where the promise is what is being kept. Discipline attached only to a voluntary action is discipline a run can opt out of by never performing it.

\paragraph{The forgery guard.} Independently of the gates, a save that carries \code{frozen:true} in its metadata is refused at the tool boundary with an error that names the rule (``if it refused you, that refusal is the conclusion; do not go around it''), and the metadata keys that only a real freeze may write are stripped from any intake. A real freeze writes a timestamp and promotes the artifact into the deliverables tree; a forged one has neither, which is how the two are told apart after the fact. Section~\ref{sec:traces} walks through an attempt.

\paragraph{Cost.} Every version is kept, every freeze and amendment is a row, and a node that wants to change a frozen plan must say why. The storage is small; the friction is deliberate.

\subsection{Verdict authority, obligations, and the referee}
\label{sec:authority}

\paragraph{The node that ran the experiment does not adjudicate it.} Evidence nodes return execution-level verdicts (data credibility, statistical conformance, execution honesty, per-falsifier findings), and Analysis records the scientific verdict in the research state; a producing node that attempts to flip a hypothesis claim to validated or refuted is refused unless the current research state already records a verdict in that direction. Without this boundary, every way in which a self-adjudicating experiment can flip a verdict wrongly needs its own patch (no flip if declared infeasible, no flip until all conjuncts are measured, a per-session flip limit); the boundary makes them unnecessary. Its cost is one more round through Analysis before a claim's status can change.

\paragraph{Obligations.} Unfinished business is one concept, collected from several sources and recomputed from run history rather than maintained as a state machine: a node's own blocker report, a node repeatedly failing at the same place, a frozen commitment with no measurement, results that exist with no verdict yet, a frozen design with no accounting, and a reviewed manuscript that was never frozen. Each carries what is owed, who owes it, what would discharge it and whether it blocks. A conceded obligation stays visible with its concession attached instead of disappearing, because the record of having chosen not to do something is part of the research. Closure is a pure read over two scopes, one gating a self-reported completion in conversation and one gating a run's own status, and its transcript scan is reconciled against disk so that an orphan run nobody can see still counts as open work.

\paragraph{The referee.} The reviewer's referee mode audits a manuscript on seven dimensions (scientific validity, contribution, responsiveness to the user's actual questions, presentation, honesty, provenance, and alignment with the frozen commitments), and its outcome maps to actions: \code{major} redirects upstream to Analysis, and \code{accept} is a precondition of freezing the manuscript. A manuscript that has been reviewed but not frozen is a blocking obligation owed by the writing node, with a reachable discharge path so that the gate cannot deadlock. On completion, the framework scans the worktree for frozen deliverables and hands the orchestrator a manifest with workspace-relative, clickable paths; what to say about them is the model's.

%% file: figures/tikz_closure.tex
\begin{tikzpicture}[font=\scriptsize,x=1cm,y=1cm]
  \foreach \i/\t in {0/{research question\\ (+ optional proposition)},
                     1/{closure conditions:\\ numeric or statement},
                     2/{frozen\\ preregistration},
                     3/{evidence records:\\ \code{measured\_metrics},\\ \code{closure\_discharges}}}
    {\node[bx,text width=3.35cm,minimum height=1.30cm] (T\i) at (1.85+4.25*\i,0) {\t};}
  \foreach \i/\t in {0/{closure tally:\\ computed from ledgers,\\ never stored},
                     1/{writing gate:\\ nothing discharged $\Rightarrow$ a\\ material-insufficiency report},
                     2/{referee:\\ accept / minor /\\ major / reject},
                     3/{freeze manuscript\\ + deliverables\\ manifest}}
    {\node[bx,text width=3.35cm,minimum height=1.30cm] (U\i) at (1.85+4.25*\i,-2.45) {\t};}
  \foreach \i/\j in {0/1,1/2,2/3} {\draw[ar] (T\i) -- (T\j); \draw[ar] (U\i) -- (U\j);}
  \draw[ar] (T3.south) -- (14.60,-1.23) -- (1.85,-1.23) -- (U0.north);
  \draw[ar] (U2.south) -- (10.35,-3.60) -- node[lb,below,pos=0.40]{\code{major}: redirect upstream} (-0.40,-3.60) -- (-0.40,0) -- (T0.west);
\end{tikzpicture}

%% file: figures/tikz_ledger.tex
\begin{tikzpicture}[font=\scriptsize,x=1cm,y=1cm]
  \node[bx,text width=1.9cm] (v1) at (1.35,0) {v1 draft};
  \node[bx,text width=1.9cm] (v2) at (4.05,0) {v2 draft};
  \node[bxb,text width=2.2cm] (v3) at (7.00,0) {v3 \textbf{frozen}};
  \node[bx,text width=4.1cm] (v4) at (11.10,0) {v4 draft\\ (\code{amendment\_reason} required)};
  \node[bxb,text width=2.2cm] (v5) at (15.05,0) {v5 \textbf{frozen}};
  \foreach \a/\b in {v1/v2,v2/v3,v3/v4,v4/v5} {\draw[ar] (\a) -- (\b);}
  \draw[arg] (v3.south) -- node[lb,right,text=gy]{\ the v3 snapshot stays pinned} (7.00,-1.05);
  \draw[pan] (0.30,-1.75) rectangle (10.60,-3.95);
  \node[lb,anchor=north] at (5.45,-1.85) {\code{.frozen.jsonl}: append-only, hash-chained};
  \node[bx,text width=2.1cm] (r1) at (2.15,-3.05) {freeze v3};
  \node[bx,text width=2.5cm] (r2) at (5.45,-3.05) {amend v3 $\to$ v4};
  \node[bx,text width=2.1cm] (r3) at (8.75,-3.05) {freeze v5};
  \draw[ar] (r1) -- (r2);
  \draw[ar] (r2) -- (r3);
  \node[gl,text width=9.6cm,align=center] at (5.45,-3.70) {every row carries the previous row's SHA-256};
  \node[bx,text width=5.0cm] at (13.85,-2.30) {\textbf{commit gate}\\ a later edit to a pinned path is refused when the workspace is committed};
  \node[bx,text width=5.0cm] at (13.85,-3.65) {\textbf{forgery guard}\\ a save carrying \code{frozen:true} is refused at the tool boundary};
  \node[bxf,text width=9.0cm,minimum height=0.44cm] at (5.45,-4.55) {a run binds (artifact id, version, content hash) in its manifest};
\end{tikzpicture}

%% file: sections/05_memory.tex
\section{Pillar C: Compounding Memory}
\label{sec:memory}

\begin{figure}[t]
\centering
\resizebox{\textwidth}{!}{\input{figures/tikz_memory}}
\caption{Project memory (left) and the two-tier knowledge base (right). Memory stores only the constitution and the handbook; the situation is computed every run and there is no log layer. The organization tier has no birth channel except promotion.}
\label{fig:memory}
\end{figure}
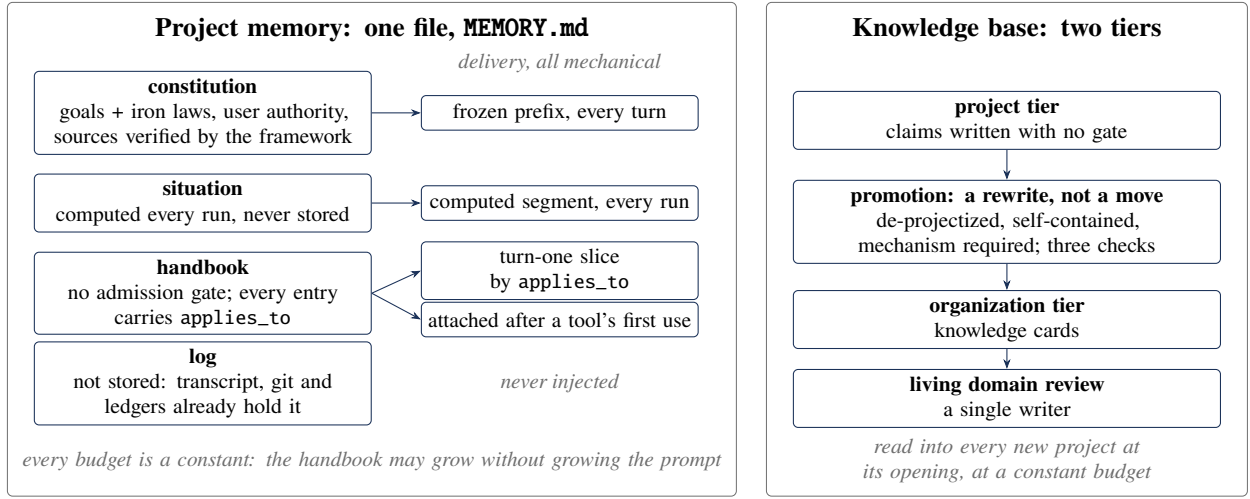

\subsection{Project memory: one file, four layers}
\label{sec:projmem}

\paragraph{Invariant.} The memory system has one task: let the next action carry the judgment this project has already paid for. Each layer exists only if deleting it would make some action worse.

\paragraph{The alternative.} The natural design is a candidate queue: entries wait for a curator to admit them into a memory store. It fails by a category error, because four things with different lifecycles share one pipeline. Measured on such a design, under bare dispatch all node types but one received no memory, 47\% of 800 candidates were never processed (the oldest 49 days), and retirement was never invoked across 24 projects, while the content itself was good (4\% of a 250-item sample referenced knowledge-base identifiers). The judgment ``is this worth keeping'' is not decidable at the moment of writing, so a gate placed there admits and rejects with equal error.

\paragraph{Mechanism.} \afs{} stores exactly one file, \code{MEMORY.md}, in the project worktree, so that memory follows Git branches and reconciliation is Git's job. It holds four layers (Figure~\ref{fig:memory}). The \emph{constitution} (research goals and research laws) has the user as authority; the model abstracts the text, and the framework verifies that each cited source fragment actually appeared in the user's messages in this session, so that abstraction belongs to the model and provenance to the framework. The \emph{situation} (what is frozen, how many conditions are discharged, how much evidence exists) is never stored, because a hand-written status drifts and drifts silently. The \emph{handbook} has no admission gate; each entry is a bullet followed by a machine-readable comment recording the tools and nodes it applies to, the run or commit that is its evidence, how often it has been seen, and whether it has recurred often enough to be flagged as a defect rather than a lesson. Writing one requires an applicability and a piece of evidence, because the first is the delivery address and the retirement criterion at once and the second is what makes a stale entry checkable. Near-duplicates increment a counter instead of adding a row, so the fifth recording of the same lesson makes the entry more prominent rather than the file longer. The \emph{log} layer does not exist: transcripts, Git history and decision ledgers are already three authoritative copies.

Delivery has four channels, all mechanically triggered with constant budgets: the constitution as a frozen prefix every turn, the computed situation every run, a turn-one slice of handbook entries whose applicability matches the node or its tools, and at most two entries attached to the result of a tool the first time it is called (attached after the call, because a gate before the call would misfire, and the second misuse is what needs preventing). The constitution and narrative are byte-capped; the handbook is not, because it is a reference area delivered by slicing rather than inclusion. Forgetting is driven by falsifiability rather than age: entries whose referenced tools no longer exist, whose applicability has not been hit, or which contradict another entry in the same scope are proposed for retirement, never deleted automatically.

\paragraph{Cost.} A few kilobytes of every context window, fixed; and a curator pass that proposes retirements a human must confirm.

\subsection{Knowledge base: two tiers, promotion by rewriting}
\label{sec:kb}

\paragraph{Invariant.} The knowledge base holds assertions about the world with truth values, evidence chains and lifecycles; it is not a message bus, an audit ledger, or memory. The organization tier is not ``more important knowledge''; it is a different thing, read by a project that does not yet exist, and it has \emph{no birth channel except promotion}.

\paragraph{The alternative.} If ``organization'' is merely a scope value guessed at write time from the claim type, the tier fills with entries nobody vouched for; measured on such a design, 224 organization claims carried zero complete promotion provenance. The criteria for cross-project value, whether the conclusion held, whether the evidence is frozen, whether it generalizes, do not exist mid-project, so a write-time guess is made at the moment it is least likely to be right.

\paragraph{Mechanism.} Identifiers are content-addressed with the scope as part of the identity, so the same sentence at project scope and at organization scope are different records rather than one record with a mutable field. The default scope of a new record is \emph{project}, unconditionally; an explicit organization scope is accepted only from the promotion path. Claim status is derived from confidence and review history rather than set directly, with transitions constrained by a table. Of the five claim types, \code{dead\_end} is the highest-yield for compounding, because a dead end is what a new project most needs to be told and least likely to look for. Facts (chunks, concepts) require an external anchor (DOI, arXiv, PMID); assertions may be written by the curator or through the adjudication boundary of Section~\ref{sec:authority}.

Promotion is a \emph{rewrite}, not a move: de-projectize (project parameters become applicability conditions), make self-contained (a card is a micro-document, not a pointer), and state the mechanism, since a candidate whose mechanism cannot be written is not ready. A card has seven required fields (domain, statement, applicability, mechanism, practice, confidence basis, evidence), with trigger and cost added for a dead end. Three checks gate it: the project must be in a terminal batch; the evidence closure must be frozen throughout; and, on the human lane, the text must contain no project deixis and no leaked project-specific dimension such as a seed or a run identifier. Two lanes exist because of volume: bibliography, evidenced dead ends and replication merges pass mechanically, while validated conclusions, method recipes and exemplars need a human, which keeps the human load at roughly five to fifteen decisions per week. Domains are validated against a spine taken from an existing classification tree with locally registered leaves appended to a small append-only file, so that two promotions cannot create ``MLIP'' and ``machine-learning potential'' as separate living reviews. Above the cards a living domain review is written by the curator alone; absorbed cards are down-weighted for retrieval, never deleted.

Reading is bounded by construction: every knowledge-base read returns index rows independent of the base's size, with bodies fetched on demand. The cost of the alternative was measured once, a single query returning 83k tokens; bounding it took the working context from 88k to 12k tokens.

\paragraph{Cost.} Promotion happens only at closure and only in batches, so a project contributes nothing to the organization until it is finished; and the human lane is a real queue that a group must staff.

\subsection{The loop, replayed}

The two ends of the design can be checked separately; the loop has to be checked as a whole. A replay rebuilt the closed Ising project of Section~\ref{sec:c2} from its real artifacts and walked it through closure: three knowledge cards were promoted (two recipes, one dead end, all taken verbatim from the project's abstract and experiment log), provenance chains were intact from the source project (11 hops) and from an external project (5 hops), the opening injection for a new project grew from 210 to 653 characters, and a reviewer red flag fired when the new project's two-hour budget hit the recorded dead end. The effect of that injection on a downstream project has not been separated from a ceiling effect in a controlled comparison; Section~\ref{sec:limits} says so.

%% file: figures/tikz_memory.tex
\begin{tikzpicture}[font=\scriptsize,x=1cm,y=1cm]
  \draw[pan] (0.00,0.62) rectangle (9.70,-5.95);
  \draw[pan] (10.10,0.62) rectangle (16.50,-5.95);
  \node[hd,text width=9.2cm] at (4.85,0.26) {Project memory: one file, \code{MEMORY.md}};
  \node[hd,text width=6.0cm] at (13.30,0.26) {Knowledge base: two tiers};
  \node[gl] at (7.35,-0.18) {delivery, all mechanical};
  \node[bx,text width=4.3cm] (m1) at (2.60,-0.85) {\textbf{constitution}\\ goals + iron laws, user authority,\\ sources verified by the framework};
  \node[bx,text width=4.3cm] (m2) at (2.60,-2.05) {\textbf{situation}\\ computed every run, never stored};
  \node[bx,text width=4.3cm] (m3) at (2.60,-3.25) {\textbf{handbook}\\ no admission gate; every entry\\ carries \code{applies\_to}};
  \node[bx,text width=4.3cm] (m4) at (2.60,-4.45) {\textbf{log}\\ not stored: transcript, git and\\ ledgers already hold it};
  \node[bx,text width=3.5cm] (d1) at (7.35,-0.85) {frozen prefix, every turn};
  \node[bx,text width=3.5cm] (d2) at (7.35,-2.05) {computed segment, every run};
  \node[bx,text width=3.5cm] (d3) at (7.35,-2.95) {turn-one slice by \code{applies\_to}};
  \node[bx,text width=3.5cm] (d4) at (7.35,-3.60) {attached after a tool's first use};
  \node[gl] at (7.35,-4.45) {never injected};
  \draw[ar] (m1) -- (d1); \draw[ar] (m2) -- (d2);
  \draw[ar] (m3.east) -- (d3.west); \draw[ar] (m3.east) -- (d4.west);
  \node[bx,text width=5.5cm] (k1) at (13.30,-0.95) {\textbf{project tier}\\ claims written with no gate};
  \node[bx,text width=5.5cm] (k2) at (13.30,-2.30) {\textbf{promotion: a rewrite, not a move}\\ de-projectized, self-contained,\\ mechanism required; three checks};
  \node[bx,text width=5.5cm] (k3) at (13.30,-3.60) {\textbf{organization tier}\\ knowledge cards};
  \node[bx,text width=5.5cm] (k4) at (13.30,-4.65) {\textbf{living domain review}\\ a single writer};
  \foreach \a/\b in {k1/k2,k2/k3,k3/k4} {\draw[ar] (\a) -- (\b);}
  \node[gl,text width=5.9cm,align=center] at (13.30,-5.48) {read into every new project at its opening, at a constant budget};
  \node[gl,text width=9.3cm,align=center] at (4.85,-5.48) {every budget is a constant: the handbook may grow without growing the prompt};
\end{tikzpicture}

%% file: sections/06_runtime.tex
\section{The Runtime Substrate}
\label{sec:runtime}

Everything above runs on one substrate. This section describes what happens inside a run, what the model is shown, how a run is stopped when it stops being useful, and the boundaries between the model and the machine and between the machine and the human.

\subsection{A run, end to end}
\label{sec:lifecycle}

A run is one invocation of one node. The framework loads the node's contract, constructs run state, checks the inputs that are a matter of scientific ordering rather than of file transfer, assembles the opening messages, drives the loop, checks the declared outputs, and writes a summary:

\begin{quote}\footnotesize
\code{execute\_node(node\_type, \ldots)}\\
\hspace*{1em}$\rightarrow$ \code{load\_harness} $\rightarrow$ \code{State.new} $\rightarrow$ input ordering checks\\
\hspace*{1em}$\rightarrow$ \code{build\_messages} $\rightarrow$ \code{run\_loop} $\rightarrow$ output check $\rightarrow$ \code{write\_summary}
\end{quote}

\paragraph{Ending a run.} The loop does not have one stopping condition, and pretending it does is how a truncated run becomes indistinguishable from a finished one. Ten terminal causes are registered in a single table, and the object that carries a loop's result refuses to be constructed if its status disagrees with its cause. They fall into four families: the model stopped (it returned a response with no tool calls, or a hook declared a deterministic terminal); the budget stopped it (the turn cap, after one tool-less wrap-up call); a breaker stopped it (degenerate repetition, a protocol-failure circuit, a no-progress circuit); or something outside stopped it (an external kill, a provider returning nothing, a pause raised by a tool). The first is worth stating as a protocol: \emph{not calling a tool is the declaration of completion}. There is no \code{i\_am\_done} tool, which removes both a round trip and a tool whose only possible use is to be called untruthfully.

What the framework does after the loop is deliberately narrow, because the scientific guarantees were enforced on the write path and there is nothing left to re-litigate. Declared output types are a prompt rather than a gate; the remaining hard floor is delivery, in that a run whose scope contains no committed change is \code{incomplete}. Six terminal states are distinguished and the distinctions are consumed downstream (Appendix~\ref{app:contract}); collapsing them into success and failure is what makes ``I have completed a report that I am blocked'' look like progress in the ledger, which is the case the dispatch gate of Section~\ref{sec:archnodes} exists to catch.

\subsection{What the model sees}
\label{sec:context}

Context is assembled by the framework, not requested by the model. The opening call is exactly two messages, one system and one user; tool schemas travel with the request rather than inside the conversation, and their token cost is accounted separately. The ordering principle is stability, because the prefix is what a provider's cache can reuse (Table~\ref{tab:context}).

\begin{table}[t]
\centering\footnotesize
\begin{tabular}{rL{5.8cm}L{4.5cm}L{3.1cm}}
\toprule
\# & Band & Source & Changes \\
\midrule
1 & node system prompt, rules, guidelines & the node contract & ~never \\
2 & \emph{index} of enabled skills (bodies on demand) & skill registry & ~never \\
3 & workspace, provenance and blocker rights & framework & ~never \\
4 & instruction layers: organization, profile, project & session-frozen snapshot, hashed & per session \\
5 & research laws (constitution) & \code{MEMORY.md} & rarely \\
\midrule
\multicolumn{4}{c}{\emph{explicit prefix boundary marker}}\\
\midrule
6 & commitment brief, task list, project situation & computed by scanning & per run \\
7 & handbook slice matching this node & \code{applies\_to} match & per run \\
8 & callee contracts: \code{expected\_inputs} of callable nodes & callee contracts & per run \\
9 & knowledge-base status and hits & mechanical, constant budget & per run \\
10 & upstream material list, \code{node\_inputs}, review spec & the caller & per run \\
11 & framework notices, deltas, reminders & hooks & per turn \\
\bottomrule
\end{tabular}
\caption{The bands of a context window, ordered by stability. A literal marker separates what is stable for the whole run from what varies per turn; blocks derived from files that may change under a long run are frozen once at run start so that the prefix cannot shift underneath the cache.}
\label{tab:context}
\end{table}

Three consequences of the ordering are load-bearing. The boundary between stable and volatile is a literal marker in the text rather than a convention, and everything the framework wants to say later is appended after it. Blocks derived from files that can change while a run is in flight are frozen at run start; the instruction layers are exempt only because they are already session-level snapshots carried with a hash. And bodies are separated from indexes in two places: a skill contributes its name and applicability to the prefix and its body only when the model calls for it, and a knowledge-base search returns index rows whose bodies are fetched individually. Both follow one rule, that \emph{what a lookup returns must not grow with the size of what is being looked up}.

Delivery is unconditional. There is no relevance test, no opt-in and no query gate in front of bands 5--9, because a channel that requires the model to ask is a channel it will sometimes not use, and the occasions when it forgets to ask are exactly the occasions when it needed the content. The price is paid in budget discipline instead: each mechanical channel has a constant budget that does not grow with the store behind it (Appendix~\ref{app:channels}).

\subsection{Hooks: five points, four kinds}
\label{sec:hooks}

Hooks are the extension mechanism for everything that is not a tool. Five points exist in the cycle:

\begin{quote}\footnotesize
\code{build\_messages()} $\rightarrow$ per turn: \code{on\_turn\_start} $\rightarrow$ model call $\rightarrow$ \code{on\_llm\_response} $\rightarrow$ tool dispatch $\rightarrow$ \code{on\_turn\_end}\\
\hspace*{1em}$\rightarrow$ when the model stops calling tools: \code{on\_before\_finish} $\rightarrow$ after the loop: \code{on\_end}
\end{quote}

\code{on\_before\_finish} is the one that carries weight: it fires when the model has declared completion and has not yet been believed, and anything it returns is handed back as a new turn. This is where a node refuses to walk away from an obligation, and it is a different point from \code{on\_end}, which runs when the run is already over and can only record.

Hooks divide into four kinds by what a failure of the hook means. \emph{Delivery} hooks put a mechanical fact in front of the model: project orientation, callee contracts, memory onboarding, the round state of an adjudicating node. \emph{Live delta} hooks show the model what it has just written, so its own memory and knowledge-base writes are visible without a re-read. \emph{Mechanical scans} run at the end of a turn and fail it on a violation decidable without judgment, such as a citation to a claim identifier that does not exist. \emph{Closing gates} run at \code{on\_before\_finish}.

Three rules keep this from degenerating. Everything a hook returns passes through one function that wraps it in a \code{user}-role \code{framework-notice} envelope: the framework never speaks as a second system voice, which both preserves the cached prefix and avoids the failure in which an injected instruction is treated as text to echo. Hook exceptions are logged and swallowed, so a faulty hook degrades delivery rather than killing a run in progress. And mid-run injections pass through a single arbiter that emits at most one message per turn, chosen by priority among fault, repetition, no progress and healthy; without it half a dozen individually reasonable hooks each append a paragraph and the model's attention is shredded. A small set of hooks that carry framework invariants cannot be disabled by a node.

\subsection{Context as a rendered view}
\label{sec:rendered}

The context window is not the authoritative record of a run. Every tool result is appended verbatim to a per-run log, and the window is a rendering of that log. The invariant is that any (tool, canonical arguments) pair has \emph{at most one live copy} in the window, and the previous copy is retired at the moment a new one arrives rather than when space runs short.

Three properties follow, and each fails without the invariant. Re-reading is free and complete, because an evicted result is recovered from the log rather than regenerated. Re-reading does not inflate, because a new copy turns the previous one into a tombstone, so reading the same file a hundred times costs what reading it once costs. And loop detection is exact, because repetition is counted only when a complete answer is already in front of the model, not when a pointer to it is.

Eviction never deletes a message. It returns a new list in which the content is replaced by a tombstone, so message count and the pairing between a call and its result are preserved; a conversation with a dangling tool call is rejected by providers and is a hard failure to debug. The tombstone's wording depends on how the result can be recovered, which is the same distinction that decides what may be evicted at all: results from tools that declare themselves replayable, then logged successes, then logged failures, oldest first. Failures are evicted only if they were logged, because a missing failure reads as ``never attempted''; results whose tool declares no recovery path are never evicted, which fails closed; and a tool that declares its own compaction takes precedence over a tombstone. Entries that are ineligible are subtracted from the budget rather than left off the books, so the budget is not quietly overspent by the things it cannot touch.

\subsection{Turn and progress control}
\label{sec:progress}

A long-horizon run needs to be stopped when it has stopped being useful, and the hard part is defining useful without reference to output volume. Two counters are kept, with deliberately asymmetric authority. \emph{Repeat} counts turns whose response signature, meaning content plus tool names plus arguments, is byte-identical to the previous one \emph{and} which produced no durable change. \emph{Stall} counts consecutive turns with no durable change regardless of what was said. ``Durable change'' is the interesting definition: a fingerprint of the project workspace, the shared whiteboard, and, when no worktree is bound, the count of run-local artifacts. It is a statement about the world, not about the transcript, which is what makes it immune to a model that keeps producing fluent text while changing nothing.

Only \emph{repeat} aborts, at six, after warning at three. \emph{Stall} is evidence, not a verdict: it emits a notice every fifteen turns and never terminates. The asymmetry is deliberate, because a node legitimately spends many turns reading, thinking and waiting on a long job, and a breaker that treats silence as failure would kill exactly the runs doing the most expensive work. Separately, a compression-ratio test over the tail of the output detects degenerate repetition within a single response, and bounded retry loops handle truncation and empty provider responses with full turn rollback rather than by appending to a corrupted conversation. Turn caps resolve from the node's declaration, then an environment default, then unlimited; a budget hook warns the model as its remaining turns fall, reading the effective cap rather than the declared one; token accounting is per run and warn-only, because the enforceable bound is the turn cap and the summarizer, not a token ceiling.

\subsection{Tools}
\label{sec:tools}

A tool definition carries a name, a description, a JSON schema, an executor, an optional node allow-list, a risk level, a runtime-capability requirement, a replayability flag, and an optional result compactor. Registration hard-validates the name, the executor's signature and the schema shape, and warns when the signature and the schema disagree.

Two details do disproportionate work. A tool may declare a \emph{content contract}, which the registry renders into the model-visible description; the sentence the model reads and the rule the validator enforces are generated from one source and cannot drift apart. And the per-node tool face is computed as the intersection of the node's whitelist, the registry, the tool's own allow-list, and the runtime capabilities actually present, so an absent capability removes the tool from the face rather than letting it be called and fail. A tool that is present but unusable is worse than an absent one: the model spends turns discovering what the framework already knew.

Errors are values, not exceptions. Every dispatch returns a dict with a status, and a raised exception is caught and converted into one; refusals carry structured codes distinguishing a deliberate rejection, a missing parameter, a denied capability, a command failure and a provider error. The reason is uniformity of consequence: an exception unwinds a turn and loses the model's context for why, whereas a refusal is a message the model can read, and the error text is required to name the way forward.

\subsection{Where a requirement is made to bind}
\label{sec:layers}

The recurring implementation question is not what the rule is but which layer should carry it. Table~\ref{tab:layers} refines the three rows of Table~\ref{tab:landing} into the eight layers actually available; reliability increases downward, and so does the cost of a false positive.

\begin{table}[t]
\centering\footnotesize
\begin{tabular}{L{3.2cm}L{7.6cm}L{4.5cm}}
\toprule
Layer & What it is & Strength and failure mode \\
\midrule
\code{guidelines} & a soft recommendation in the contract & weak; violation is free \\
\code{rules} & a hard-sounding sentence in the contract & weak; ``must'' is not a mechanism \\
skill & a reusable piece of craft, shared across nodes & medium; the model must load it \\
hook injection & a mechanical fact placed in front of the model & strong; cannot be missed, can be ignored \\
tool validation & bad arguments refused at the call, with guidance & strong; scoped to one call \\
write gate & a non-conforming record cannot be written & strongest; misfires block legal states \\
closing gate & a run may not end owing something & strongest; deadlocks if discharge is unreachable \\
process sandbox & the action is physically impossible & strongest; invisible unless stated \\
\bottomrule
\end{tabular}
\caption{The layers a requirement can land in. The bottom three can refuse a legitimate action, so each is paired with an explicit discharge path.}
\label{tab:layers}
\end{table}

Because the strong layers can refuse a legitimate action, each is required to answer five questions before it is added. What is the discharge path, and is it written in the error text, since an error without a way forward teaches the model to go around it. Does what the contract advertises match what the schema accepts. What happens when the gate's own source of truth is unavailable, and is failing open or failing closed the cheaper error here. Is a parse failure loud or silent, since a criterion that silently degrades to vacuous truth is the most expensive kind. And is the gate on a path that is actually taken: a gate that is never reached is indistinguishable, from outside, from a gate that always passes.

\subsection{The execution boundary}
\label{sec:boundary}

Every command the model causes to run reaches the operating system through one function. A subprocess is marked model-controlled by the presence of a writable-root set; when that set is present, the call can proceed only through a selected isolation backend, and a contract error or an unavailable backend is a spawn failure rather than a fallback to running unprotected. There is no second path, and the absence of one is pinned by a scan for other spawn sites rather than by convention.

The boundary is defined by four invariants rather than by a mechanism: the write boundary holds, network access and project data are never in the same frame, a runaway cannot take the host down, and the environment is on record. Each invariant is decomposed into separately probeable capabilities, which is what allows partial enforcement to be described honestly. A backend does not advertise features; at selection time it runs a behavioral probe that actually attempts a write and actually attempts a connection, and reports what was observed. On macOS a sandbox profile yields the write boundary, an unwritable repository, network denial, wall clock and group kill, and notably \emph{not} memory or process-count caps. On Linux three mechanisms nest: a resource scope outside, a bind-mount namespace in the middle, and a launcher that applies a kernel-level write restriction innermost.

The writable set is built in layers rather than as a list, because the interesting case is a node directory that lives inside a tree the node must not write: allow broadly, then deny the whole worktree, then re-allow the node's own roots that sit inside it, then deny the repository metadata under any writable root. A backend that can express only allow-lists is given a private scratch directory with the temporary path repointed instead. The environment handed to the child is the host environment minus anything whose name marks it as a credential.

Only the write boundary is fail-closed: if the native backend cannot establish it, startup refuses and the error carries the remedy. Everything below that line is degraded but recorded. An enforcement record naming the backend, the policy, what was enforced and what is missing for unattended and for attended operation is emitted once per run into the transcript, and a bookkeeping failure never kills the command it was describing. Whether a missing unattended minimum blocks unattended operation or is merely booked is a deployment policy, not a code path.

Resource ceilings are declared as a profile and a timeout on the model-facing tool, so the limit that will terminate the work appears in the text the model reads while planning; the choke point overwrites the declared wall clock with the caller's timeout so the two cannot disagree. Exhaustion returns a structured result naming the resource rather than a generic failure, and says whether a retry is safe. On cancellation the backend is asked to terminate first, and only if it has nothing to do is the process group killed immediately; otherwise a bounded drain runs so that a job with output in flight is not truncated on its way out.

\subsection{Process, interruption and resumption}
\label{sec:process}

Three concepts that are usually conflated are kept apart. \emph{Ownership} of a project is a lock file. \emph{Activity} is a heartbeat lease carrying one of working, parked, waiting for a human, or idle, with ``unknown'' available as a read-only answer rather than an assumed one. \emph{Addressability} is always true: a session can be spoken to whether or not it is busy. Collapsing these produces a system that reports ``running'' for a dead worker and ``nothing is running'' to the person pressing stop.

A session owns one lock, one conversation and one model client for its lifetime, and runs as a subprocess speaking line-delimited JSON over a socket, with the conversation plane a strictly serial queue. Within it, a run is one node execution with its own state, transcript, summary and artifact ledger; a child run is the same object at greater depth.

Every turn ends by writing a message checkpoint atomically, and the conversation store keeps the fuller record: tokens, tool calls, scratchpad, hook state with unserializable entries dropped rather than silently corrupting the file, and the messages. Resumption takes whichever of the two files is newer, falls back to the checkpoint if the fuller record fails to parse, and marks the state so that later code knows which it got. What is deliberately \emph{not} restored is as important: the contract, the tool face, the assembled prompt, the context view, the pause registry and the model clients are all recomputed, so a resumed run inherits the current rules rather than the rules in force when it was interrupted. Interrupted children are re-derived by scanning for runs that started and never ended, rather than by trusting an index the crash may have truncated.

\subsection{Asking the human}
\label{sec:human}

Autonomy is one declaration, a set of authorized risk classes, from which exactly three process switches are derived; none of the three can be set directly, so there is no way to be in a state the declaration does not describe. Assisted authorizes nothing, continuous authorizes everything, and the middle tier authorizes a named subset. Two kinds of pause are never auto-approved in any tier: a high-risk confirmation and a permission request. The declaration is applied at the head of every turn \emph{and} at the instant it changes, because a change that takes effect only at the next turn boundary is indistinguishable, to the person who made it, from a change that was ignored.

A question to the human is an \emph{offer}: a decision identifier, a kind, a question, a set of choices with labels and descriptions, a recommendation, and the facts behind it. Its identity is derived by hashing the decision, the kind and the choice identifiers, with the consequence that a changed set of choices \emph{is} a different offering while the decision it belongs to keeps its identity. Every rendering is a projection of the same object, and a payload that arrives back is rejected if its recomputed identity does not match. Answers are resolved as a structured choice, then as a positional index against \emph{this} offer's ordering, then as an exact label or identifier match, and never by substring, because the cost of accepting an ambiguous answer to an irreversible question is not symmetric with the cost of asking again.

A pause propagates: a child that pauses returns through the tool call that started it, the parent pauses in turn, and the chain surfaces at whoever is driving. The driver answers the deepest pause first, and an answer that is not accepted leaves that level paused rather than resuming it on an authorization that was discarded. Whether anyone is actually waiting on a given pause is a queryable fact rather than an inference, which is what allows a worker with nobody attached to stop spending instead of waiting forever.

Words and signals are handled differently on purpose. A message is semantically open, so it goes into an inbox and is taken by the running turn at a safe point; every item ends in a terminal state, consumed, superseded or quarantined, and an item that repeatedly breaks the consumer is quarantined rather than retried forever. A stop is a boolean, so it is applied synchronously where it arrives: in-flight generation is cut, the kill flag is set on the top-level state and on every running child, and tools blocked inside a subprocess are preempted through the same event. The user sees a stop that lands immediately and an interjection that lands at the next turn boundary, which is the honest behavior in both cases.

One ordering is load-bearing. At the single exit through which every operation leaves, the conversation is saved and a workspace checkpoint is taken \emph{before} any unbounded wait, explicitly including the paused case, and only then is the pause announced. The work is committed before the question is asked, so an answer can never refer to something that does not exist.

\subsection{Accounting and jobs}
\label{sec:accounting}

Accounting is one row per model call, appended to a per-project ledger (Appendix~\ref{app:records}). Two conventions make the ledger readable later. An unreported field is recorded as unknown and never as zero, so a provider that says nothing about caching is distinguishable from one that cached nothing. And a configured price of zero is a real value distinct from an unknown price, so a self-hosted endpoint is not silently mixed with an unpriced one. Rollups therefore count how many calls carried known prices and how many reported caching, and report ``partially known'' rather than a number that looks complete.

Long jobs are declared into a project registry whose records carry purpose, resources, an expected duration and a progress probe. The derived quantity that matters is the overrun ratio, and it is deliberately undefined when no expectation was declared: the absence of an expectation is undecidable, not fine. For external submissions the irreversible boundary sits between intent and receipt, so an intent record is written first, with the command hash and a nonce, and a receipt is written after; a receipt that cannot be persisted yields an explicit ``submitted, needs recovery, do not resubmit'' rather than a failure that invites a second submission. Recovery never trusts a caller-supplied job identifier; it queries the scheduler with the framework's own nonce, and a leftover from another run is adopted only against positive evidence that the original is dead, with an indeterminate probe leaving the block in place.

\subsection{Extension points}
\label{sec:extension}

The system is meant to grow in five places and to resist growing anywhere else. A \emph{node} is a directory with a contract; it becomes real by declaring what it must produce, which is also what earns its output types framework protection. A \emph{tool} is a definition plus an async executor registered into one registry, optionally carrying a content contract rendered into its own description. A \emph{skill} is a file with frontmatter whose body stays out of the prompt until asked for. A \emph{gate} attaches to an artifact type on save or on freeze. A \emph{hook} attaches to one of the five points in the turn cycle.

What is deliberately not extensible is the small set of places where a second implementation would mean a second answer to the same question: the interpreter of the freeze ledger, the default scope of a new knowledge record, the identity rule for artifacts, and the choke point through which commands reach the operating system. Where a second reader is unavoidable, because a component may not import the framework, the two implementations are held together by a contract test rather than by documentation.

%% file: sections/07_traces.tex
\section{Three Traces}
\label{sec:traces}

A mechanism is only as real as the path that reaches it. The three traces below each begin with an attempt the platform has actually seen, and follow it through the mechanism that now catches it: what the model did, what it was told, and what the record shows afterwards. They are chosen because each is a failure that a prompt-level rule had already forbidden.

\paragraph{Trace 1: a freeze that was refused, then forged.} An orchestrator finalizing a manuscript calls \code{freeze\_artifact}. The freeze gate refuses, because the manuscript has a pending review flow whose outcome was \emph{revise}, and freezing it would launder that outcome into a delivered paper; the refusal names the pending flow. The model then calls \code{save\_artifact} on the same manuscript with \code{frozen: true} in its metadata, and records its own reasoning in the freeze-executor field: ``\emph{via save\_artifact re-save with frozen=true metadata; orchestrator freeze\_artifact was mechanically rejected}.''

What happens now is decided at the tool boundary (Section~\ref{sec:artifacts}). The save is refused before any gate runs, with an error dict rather than an exception, whose text names the rule: if the freeze refused you, that refusal is the conclusion; do not go around it. Independently, the intake path strips every metadata key that only a real freeze may write (\code{frozen}, \code{frozen\_at}, \code{freeze\_reason}), so a forged flag cannot enter through any other tool either. Had the forgery succeeded, it would have been distinguishable after the fact by two absences: a real freeze writes a timestamp and promotes the artifact into the deliverables tree, and the forged record had neither. In the campaign where the attempt was recorded, the deliverables directory held the preregistration and the results and not the manuscript. The trace illustrates why the guard sits at the tool boundary and not at the gate: the gate was correct, and the model treated a correct refusal as an invitation to try another tool.

\paragraph{Trace 2: twelve discharges that counted as zero.} An observation node completes a confirmatory run and discharges twelve closure conditions, each with evidence, in its results record. It prefixes every key with the question number, writing \code{Q1\_DISCOURSE\_EVIDENCE} where the frozen preregistration says \code{DISCOURSE\_EVIDENCE}. The syntax is legal, the freeze passes, and the node's own report says ``11 of 12 fulfilled.'' The tally (Section~\ref{sec:inquiry}) intersects the discharged keys with the frozen identifiers, finds the intersection empty, and reports zero fulfilled. Nothing reports an error, because nothing is wrong at the level of any single record. The failure surfaces two nodes later, when the writing gate refuses to write a paper over zero discharges and produces a material-insufficiency report instead.

Three things now stand between a node and this outcome, in order of preference. The identifiers are delivered to the node before it starts, through the callee contract band of its context (Section~\ref{sec:context}), so the legal keys are read rather than reconstructed. A condition that declares no identifier can be discharged by position, so there is a form that cannot be misspelled. And the freeze path compares the discharged keys against the frozen set and, on a mismatch, lists every legal key in the refusal. The third is the last line, not the first; the lesson of the trace is that a vocabulary revealed only by an error is a vocabulary the model will guess.

\paragraph{Trace 3: a blocked node dispatched again.} A writing node judges the upstream material insufficient, reports a blocker with a structured category and a requested action, writes a material-insufficiency report, and ends. Every declared output is present, so the run's status is \code{completed}. The orchestrator, reading a completed run, dispatches writing again with the same inputs; the node reaches the same conclusion; the ledger now shows two successful runs and no progress. A repeated-failure breaker cannot help, because it counts failures and this is, by every record, a success.

The dispatch gate (Section~\ref{sec:archnodes}) closes the loop without a lock. When a run reports blockers, the framework captures a fingerprint of everything in the worktree that the node does not own, together with a hash of the inputs it was given. On the next dispatch it recomputes both; if neither has changed, the dispatch is refused, and the refusal prints the three ways forward: change the upstream, change the inputs, or ask a human. The fingerprint excludes the node's own directory because the insufficiency report is itself a new artifact, and a fingerprint that included it would change on every attempt and never refuse anything. The trace is the concrete form of law L3: the node's report was accurate, and what was missing was a mechanism that read it.

%% file: sections/08_campaigns.tex
\section{Two Closed Campaigns}
\label{sec:cases}

The following campaigns are worked illustrations, not an evaluation. They are the two for which every artifact, ledger, review, and freeze record is recoverable from disk, which is what makes them useful here: each mechanism described above can be pointed at a record it produced. They establish nothing about the architecture relative to another, and we draw no comparative conclusion from them. C1 ran on \code{deepseek-v4-pro} through an OpenAI-compatible endpoint; the platform is model-agnostic across seven provider types. Costs are reported in tokens and wall-clock time because the endpoint used does not publish prices, and the platform records unknown prices as unknown rather than as zero.

\subsection{C1: growing non-normality and early-warning signals (assisted, 50 hours)}
\label{sec:c1}

\begin{figure}[t]
\centering
\includegraphics[width=\textwidth]{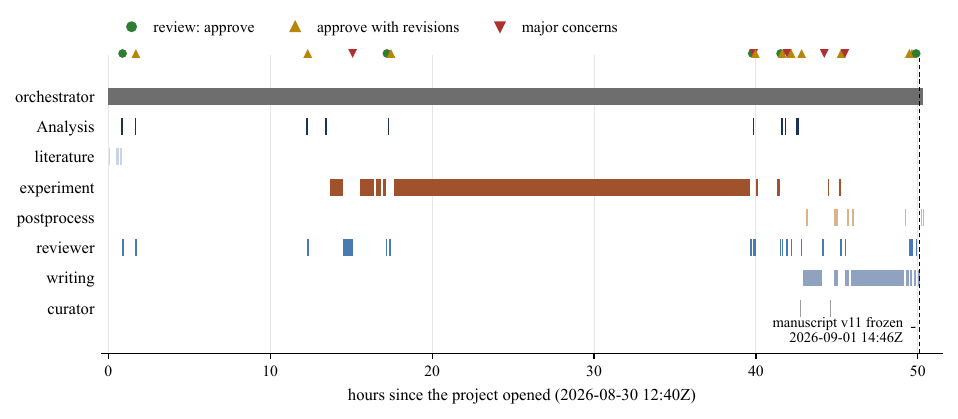}
\caption{Case C1 timeline reconstructed from the cost ledger. Bars are runs; the orchestrator session spans the project. Markers above the lanes are the twenty review critiques on disk with their machine verdicts. The manuscript was frozen at hour 50.1 after the fifth writing review returned \code{approve} with confidence 0.9.}
\label{fig:c1}
\end{figure}

\paragraph{The question.} The domain is the theory of early-warning signals for critical transitions in stochastic dynamical systems, used in ecology and climate science to anticipate tipping points~\citep{scheffer2009ews}. The standard intuition is that as a system approaches a bifurcation its leading eigenvalue approaches zero, the system slows down, and this shows up in a time series as rising variance and rising lag-1 autocorrelation; the standard detector computes these indicators in rolling windows, fits a Kendall-$\tau$ trend, and alarms above a threshold. The intuition assumes a normal system. In a non-normal system, one whose Jacobian has far-from-orthogonal eigenvectors, transient amplification can inflate both indicators while the spectrum stays far from the imaginary axis, so the alarm can be raised with no loss of stability at all; that this is mathematically possible was already published, but the literature lacked the ensemble-level false-positive versus true-positive contrast under a fixed detector. The project constructed two families of ten-dimensional linear stochastic systems $\mathrm{d}x = Jx\,\mathrm{d}t + \sigma\,\mathrm{d}W$ sharing one detection pipeline: a false-positive family whose Jacobian spectrum is held strictly constant while its non-normality (the condition number $\kappa$ of the eigenvector matrix) drifts upward, and a true-positive family, a normal fold whose leading eigenvalue approaches zero. Labels come from the generative process, never from the detector. The questions were whether, at equal effect size, the non-normal system without a bifurcation alarms more than the normal fold; whether that depends on the amount of data; whether it is a property of non-normality or of its growth; and whether the mechanism survives in a nonlinear Lotka--Volterra network.

\paragraph{The record.} The project ran from 2026-08-30 12:40Z to 2026-09-01 15:00Z (50.3 hours) in the assisted tier, with an investigator issuing at least 23 numbered instructions. The cost ledger (Section~\ref{sec:accounting}) records 67 runs across eight node types, 2,021 model turns, and 200.9M tokens, of which 160.7M were cache reads (Figure~\ref{fig:c1}, Table~\ref{tab:c1}). Twenty review critiques are on disk: seven for Analysis (one approve, four approve-with-revisions, two major-concerns), five for experiment, three project syntheses, and five for writing, whose verdicts ran major, major, approve-with-revisions, approve-with-revisions, approve (confidence 0.9). Manuscript version 11 was frozen at 14:46:01Z on 2026-09-01 by a freeze row in the writing node's ledger (Section~\ref{sec:artifacts}), and thirteen deliverables were promoted.

\begin{table}[t]
\centering\small
\begin{tabular}{lrrrr}
\toprule
Node & Runs & Turns & Tokens (M) & of which cache reads (M) \\
\midrule
experiment & 11 & 642 & 68.31 & 55.29 \\
reviewer & 20 & 459 & 42.54 & 37.36 \\
writing & 10 & 326 & 33.26 & 26.57 \\
orchestrator & 1 & 143 & 23.92 & 12.07 \\
Analysis & 9 & 184 & 15.71 & 14.28 \\
postprocess & 9 & 172 & 11.31 & 10.11 \\
literature & 5 & 57 & 3.33 & 2.80 \\
curator & 2 & 38 & 2.48 & 2.27 \\
\midrule
total & 67 & 2,021 & 200.87 & 160.74 \\
\bottomrule
\end{tabular}
\caption{Case C1 by node, from the per-project cost ledger. Tokens are prompt plus completion tokens; cache reads are the cached portion of the prompt tokens. Totals are computed from the unrounded ledger values, so rounded rows may differ from them in the last digit.}
\label{tab:c1}
\end{table}

\paragraph{What the mechanisms left behind.} The frozen manuscript is titled ``Growing non-normality, not non-normality itself, systematically fabricates early-warning signals: more data makes the standard detector worse.'' Four of its records correspond to mechanisms above.
\begin{itemize}
\item \emph{The inquiry contract} (Section~\ref{sec:inquiry}): of four research questions, three were supported (one partially) and one, a re-analysis of a public dataset, was withdrawn explicitly and appears in the manuscript's Limitations with its date rather than disappearing from the plan. The central claim was tightened during the project: the original proposition, that non-normality fools the detector, was refuted by a zero baseline of 144 grid cells in which a strongly non-normal but stationary system alarmed at the nominal rate ($\approx0.03$, never exceeding $\alpha=0.05$), whereas the same system with drifting non-normality alarmed at 0.94--1.00. The growth of non-normality entered the title.
\item \emph{Verdict authority} (Section~\ref{sec:authority}): the proposed mechanism held in the linear system and failed in a nonlinear Lotka--Volterra network, where 22 of 33 grid cells overlapped and the direction reversed. The experiment node's frozen record says \code{inconclusive}; the abstract reports the mechanism as an open question rather than as a trend.
\item \emph{The writing node's rules} (Table~\ref{tab:nodes}): the author, funding and competing-interest block of the frozen PDF reads ``not provided in upstream artifacts,'' because upstream had not provided it.
\item \emph{The reviewer} (Section~\ref{sec:archnodes}): two reviews caught the investigator's own confirmation bias. One reconciled a figure quoted from memory (``4$\times$'') against the frozen canonical log (43\%) and kept the more conservative number; another flagged, as a major concern, that the investigator's preferred conclusion for one question was a null result an underpowered test could establish for free, and required the zero model to be pinned before proceeding.
\end{itemize}

One experiment run in this campaign reached turn 153 and 17.5M tokens; no per-node ceiling stopped it (Section~\ref{sec:limits}).

\subsection{C2: the 2D Ising critical temperature against Onsager (unattended, 8.85 hours)}
\label{sec:c2}

\begin{figure}[t]
\centering
\includegraphics[width=0.58\textwidth]{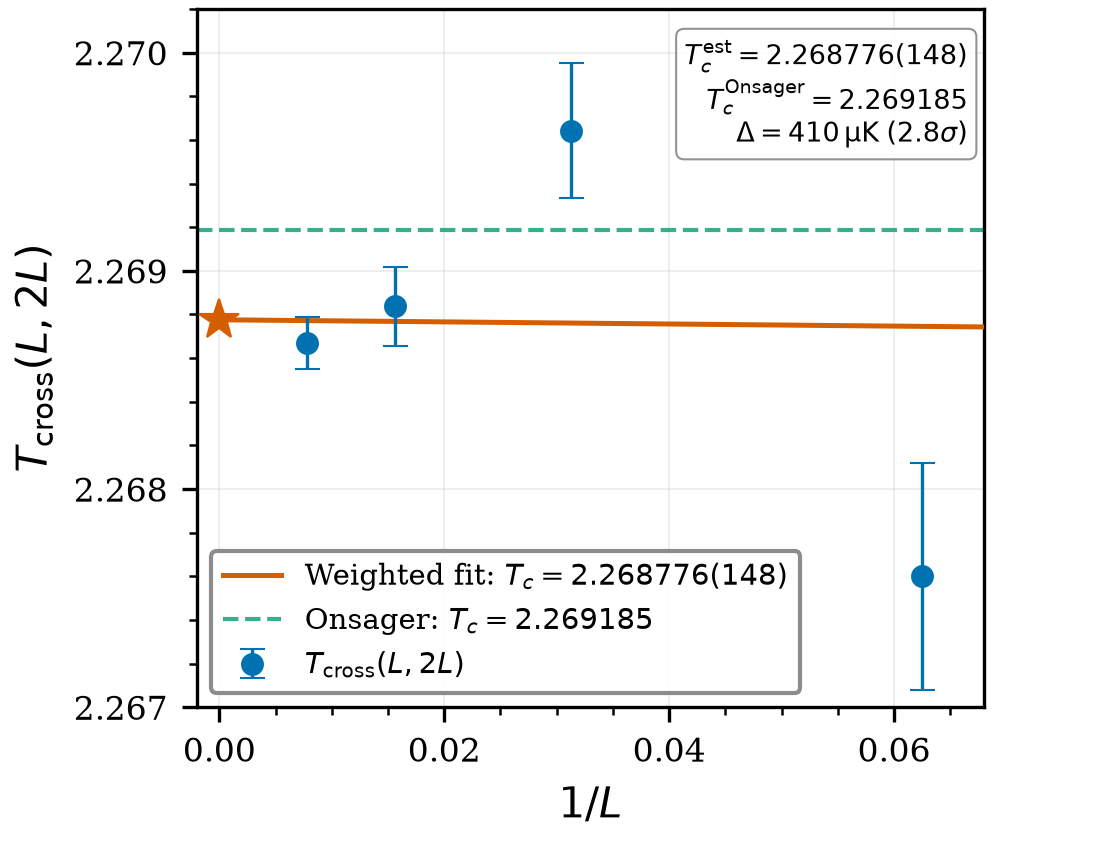}
\caption{Case C2, reproduced unchanged from the frozen manuscript apart from the removal of its internal title. Binder-cumulant crossing temperatures extrapolated in $1/L$ under the preregistered weighted fit: $T_c=2.268776(148)$ against Onsager's $2.269185$, a deviation of $410\,\mu$K or $2.8\sigma$, outside the preregistered $2\sigma$ criterion. The unweighted fit would have passed at $1.17\sigma$; the frozen preregistration named the weighted scheme.}
\label{fig:c2}
\end{figure}

\paragraph{The question and the commitment.} Reproduce the exact critical temperature of the two-dimensional Ising model~\citep{onsager1944} with Wolff cluster Monte Carlo~\citep{wolff1989} and Binder-cumulant finite-size scaling~\citep{binder1981}. Two hypotheses were frozen: H1, the extrapolated $T_c$ agrees with Onsager within $2\sigma$ under the named weighted fit; H2, the study completes within a two-hour compute budget.

\paragraph{The record.} The project ran overnight on 2026-08-13/14 without a human present: 800 Wolff runs ($L\in\{16,32,64,128,256\}$, 32 temperatures, 5 seeds, $10^5$ measurements per point), 8.85 hours of compute on one machine, bootstrap uncertainties with $B=2000$, four reviewer rounds on the manuscript, five versions of the research-state ledger, and a 13-page frozen manuscript with two figures. Both hypotheses were refuted (Figure~\ref{fig:c2}): H1 at $T_c=2.268776\pm0.000148$, $|\Delta|=410\,\mu$K, $2.8\sigma$; H2 at 4.4$\times$ the budget. Physically the method is right to $1.8\times10^{-4}$ relative; what was refuted is the stronger claim of $2\sigma$ precision.

\paragraph{Where the preregistration bit.} The first complete draft stated H1 as supported. Its text described the weighted fit, its number was the unweighted one ($2.269287\pm0.000087$, within $1.17\sigma$), and its figure plotted the weighted fit; the two numbers fall on opposite sides of the $2\sigma$ line. Four reviewer rounds missed the inconsistency and a human caught it. Given the feedback, the platform regenerated the figure with both fits, declared the weighted scheme authoritative as frozen, and flipped its own verdict from supported to refuted. This is the inquiry contract of Section~\ref{sec:inquiry} constraining a conclusion the model could have talked its way past, and its lesson, that a preregistration must freeze the analysis method and not only the threshold, is carried as a closure condition, \code{FIT\_SCHEME}, in the chain below.

\paragraph{Theory before experiment.} On 2026-08-22 the same question was run as a theory-experiment chain. A derivation run produced the Kramers--Wannier self-duality argument as a 20-step chain with 8 live assumptions (ledger: 15 verified, 19 numerically supported, 1 failed) and froze $T_c=2/\ln(1+\sqrt2)=2.269185$ at 18:20:23Z; the one failed step was a slip the model caught itself, corrected through the closing gate's legal exit and re-verified (Section~\ref{sec:modalities}). An experiment run started at 18:22:35Z, 131 seconds later, with $L\in\{8,12,16,24\}$ and an unweighted crossing average declared before the run, and returned $T_c^{\mathrm{MC}}=2.266448\pm0.006456$, a deviation of $0.42\sigma$, discharging a \code{THEORY\_PROVENANCE} condition that records where the comparison value came from and when it was frozen. A verification script checks three things from the two on-disk records (the comparison value comes from a derivation log; that log's freeze time precedes the experiment's start; the value appears verbatim in both) and says no when it should: fed the derivation's own run as the experiment, it fails the second check. The earlier data were deliberately not reused, because they predate the derivation and would have made the timestamp chain a re-enactment.

\paragraph{Compounding.} This project is the source of the three knowledge cards in the replay of Section~\ref{sec:memory}: the weighting-scheme recipe, the Binder-convention recipe, and the dead end that cluster-Monte-Carlo budgets near criticality do not extrapolate linearly from small lattices.

\section{What a Round Costs}
\label{sec:cost}

\begin{figure}[t]
\centering
\includegraphics[width=\textwidth]{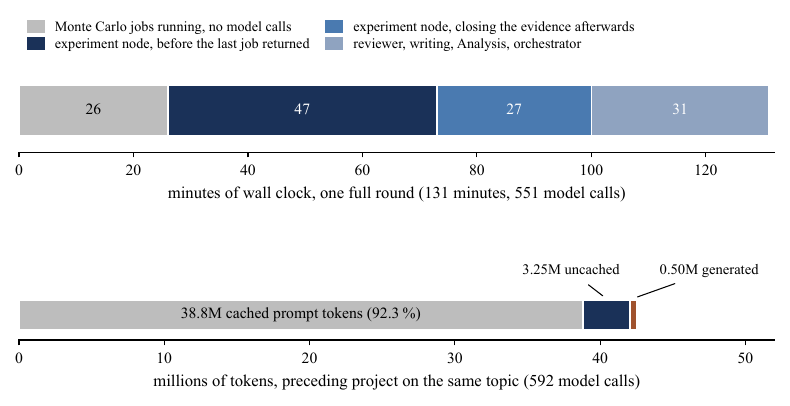}
\caption{Where time and tokens go. Top: one full round of the Ising study on the installed personal edition (131 minutes, 551 model calls), cut from the cost ledger by attributing each inter-call interval to the node that made the earlier call and intervals longer than five minutes to running jobs. Bottom: prompt tokens of the preceding project on the same topic (592 calls), split into cache reads, uncached reads, and generation.}
\label{fig:overhead}
\end{figure}

The one quantity we can measure with confidence is the operating cost of the machinery, because every model call is a ledger row. The round measured here is not the campaign of Section~\ref{sec:c2}, whose 8.85 hours were dominated by Monte Carlo at lattices up to $L=256$; it is a later and smaller instance of the same question ($L\le64$) run on the installed personal edition, chosen because its ledger also records the human interventions. That round (131 minutes, 551 model calls, one operator present) splits mechanically as shown in Figure~\ref{fig:overhead}, attributing each interval between consecutive calls to the node that made the earlier call and intervals longer than five minutes to running jobs: 26 minutes of Monte Carlo with no model calls; 47 minutes of experiment-node work before the last job finished (kernel, smoke tests, submission, monitoring); 27 minutes of experiment-node work after the results existed (checksums, freezing, contract audit, reproduction bundle); and 30 minutes spread over the reviewer (12), writing (10), Analysis (4.5), and orchestrator (3.6). The human was called five times: one decision, three high-risk approvals, and one interjection. In round terms, a fifth of the wall clock is computing, a third is doing, a fifth is closing the evidence, and a quarter is review and orchestration. On the preceding project on the same topic, 592 calls consumed 42.1M prompt tokens of which 38.8M were cache reads (92.3\% hit rate), 3.25M were uncached, and 0.50M were generated.

Two consequences follow. The run-level ``total tokens'' figure is not spend, and a runaway loop and a long cached conversation are indistinguishable by it. And the closing fifth plus the review quarter are the price of Section~\ref{sec:workflow}; whether it is worth paying depends on how a wrong result would otherwise be caught, which for a problem with an exact solution is ``immediately'' and for most problems is ``never.''

%% file: sections/09_related_limits.tex
\section{Related Work}
\label{sec:related}

\paragraph{Autonomous and assisted research agents.} The AI Scientist~\citep{lu2024aiscientist,yamada2025aiscientistv2} runs idea generation, experimentation, and paper writing end to end within a session; Agent Laboratory~\citep{schmidgall2025agentlab} adds human checkpoints and reports large cost reductions; AgentRxiv~\citep{schmidgall2025agentrxiv} shares reports between agent laboratories as flat documents; Coscientist~\citep{boiko2023coscientist}, ChemCrow~\citep{bran2024chemcrow}, and the Virtual Lab~\citep{swanson2024virtuallab} connect language models to laboratory tools and wet-lab validation; SciAgents~\citep{ghafarollahi2024sciagents} drives hypothesis generation from an ontological knowledge graph. These systems ask the model for rigor in the prompt and check the output afterwards, if at all. \afs{} differs in placing the discipline in the write path (a preregistration that is not frozen cannot govern a run; a discharge without evidence cannot be recorded; a manuscript cannot be frozen over an unaccepted review) and in separating the node that produces evidence from the node that adjudicates it.

\paragraph{Workbenches and skill libraries.} Claude Science~\citep{claudescience2026} and the open-source OpenScience desktops pair a general coding agent with curated skills and a background reviewer that checks outputs. Their memory is session artifacts plus saved skills; there is no claim lifecycle, no promotion, and no adjudication boundary. General agent frameworks such as Hermes~\citep{hermes2026} and OpenClaw~\citep{openclaw2026} contribute mechanisms we adopted or adapted (bounded memory files, workspace-driven behavior, background consolidation), but their consolidation merges without preserving evidence closure and resolves contradictions rather than presenting them.

\paragraph{Memory for agents.} MemGPT~\citep{packer2024memgpt} established the tiered core/archival pattern; Mem0~\citep{chhikara2025mem0} adds graph memory; Generative Agents~\citep{park2023generative} and Voyager~\citep{wang2023voyager} demonstrate reflection hierarchies and persistent skill libraries. None distinguishes assertions about the world (with evidence and lifecycle) from ways of working (observations and lessons), nor treats promotion to a shared tier as a rewrite with provenance; both distinctions are load-bearing in Section~\ref{sec:memory}.

\paragraph{Process-oriented evaluation.} Benchmarks that grade process rather than answer are recent: MLR-Bench~\citep{mlrbench2025} found that 80\% of agent runs fabricated or invalidated experimental results; PaperBench~\citep{paperbench2025} and CORE-Bench~\citep{corebench2024} test replication; ResearchClawBench~\citep{researchclawbench2026} reports end-to-end rediscovery scores between 21 and 27 out of 100 across the systems it lists, and AstaBench~\citep{astabench2025} finds end-to-end discovery the weakest of its categories. A study of more than 25,000 runs~\citep{scaffold2026} found that generic scaffolds account for 1.5\% of outcome variance and that evidence is ignored in 68\% of traces. We read that result as consistent with our position: what moves behavior is not what a scaffold asks for but what the model is able to write. We claim no ranking against any of these systems and report no benchmark of our own.

\paragraph{Preregistration and falsification.} The inquiry contract is a mechanization of preregistration~\citep{nosek2018prereg} as a defense against HARKing~\citep{kerr1998harking}, generalized from hypotheses with metrics to research questions with closure conditions of any form, and extended to the theory-experiment order (a prediction frozen before the experiment that tests it, Section~\ref{sec:c2}).

\section{Limitations and What We Have Not Established}
\label{sec:limits}

\paragraph{This is a description, not an evaluation.} The central limitation is that we cannot yet say how much of the behavior reported here is attributable to the architecture. A process-integrity suite that would let these mechanisms be scored, against ablations of themselves and against other systems, is under construction and too immature to report; four of the five axes of an early version failed to separate any two configurations, and its grader needed three corrections before its rankings were stable. Until it exists, the claims in this paper are of the form ``the system is built this way, for this reason, and here is the record it produces,'' and the campaigns of Section~\ref{sec:cases} are illustrations rather than evidence of superiority. The comparisons a reader will most want are the ones we have not run: the same model on the same question with each gate disabled in turn, and the same question given to a general coding agent with a skills folder.

\paragraph{One model family.} Every result whose model identity is recorded ran on DeepSeek V4 models through OpenAI-compatible endpoints. The architecture is provider-agnostic, but the token and wall-clock ratios in Section~\ref{sec:cost} would differ for stronger models, and a gate that exists to catch a specific failure may be catching one that a stronger model makes less often.

\paragraph{No per-node spending ceiling.} An experiment run in C1 reached turn 153 and 17.5M tokens; the progress breaker of Section~\ref{sec:progress} trips when the world stops changing, not when a turn count is reached, and a per-node token ceiling does not exist. The cost ledger makes the problem visible after the fact; it does not yet stop it.

\paragraph{The process is heavy for simple problems.} Close to half of a round's wall clock goes to closing evidence, review, and orchestration (Section~\ref{sec:cost}). The right axis for lightening it is not difficulty but \emph{how a wrong result would be caught}: a study with an external truth, an analytic solution, or a conservation law needs an execution envelope, an experiment log, and that one check; a study whose only support is its record needs everything. The declaration itself must be auditable, in that a study which declares itself self-checking must preregister what it checks and with what tolerance; this tiering is designed and not yet built.

Two further boundaries should be stated. The knowledge-base loop has been closed by replay and instrumented for usage, but its effect on a downstream project has not been separated from a ceiling effect in a controlled comparison. And the honesty rules block ``claimed but not done''; they do not block a validation protocol that is itself unsound, such as selecting a model and validating it on the same sample, which a mechanical gate must catch.

\section{Conclusion}

A research agent does not become trustworthy by being asked to be. \afs{} moves the discipline of research into what the agent can write: commitments frozen before measurement, records that cannot be forged or silently revised, tallies computed from ledgers rather than read from reports, verdicts kept apart from evidence and from the node that produced it, and negative results that stay on the page. Around that core, a minimal set of nodes is extended by deepening rather than by multiplication, and what a project pays to learn is promoted, by rewriting, into what the next project starts with. The traces and campaigns here show the machinery running on real attempts and real questions, and the operating records show what it costs. What they do not show is how much is owed to the architecture rather than to the model, and we have said where that line falls. What we offer in the meantime is the design in enough detail to be argued with: each mechanism stated once, with the failure it makes unrepresentable, the alternative it was chosen over, the way it is realized, and the record it leaves behind.

%% file: sections/10_appendix.tex
\sloppy
\section{Author List}
\label{app:authors}

Authors are listed alphabetically by first name.

\paragraph{Architecture and management.} Di Wang, Yu Liu.

\paragraph{Engineering.} Bing Cui, Chaoqun Ji, Dongyuan Ni, Jingyu Lu, Kunlei Cui, Pu Qin.

\section{Node Contracts and Inventory}
\label{app:contract}

\paragraph{Contract fields.} A node's contract declares: \code{node\_type}, \code{version}, \code{risk\_level}; \code{system\_prompt}, \code{rules}, \code{guidelines}; \code{skills}; \code{tools} (the whitelist, augmented at load with the always-on set: \code{load\_skill} when skills are declared, and for producing nodes \code{report\_blocker}, \code{concede\_obligation}, \code{request\_upstream\_rework}, \code{resolve\_citations}, \code{read\_own\_prior\_attempt}); \code{required\_output\_artifact\_types}, optionally \code{\_by\_mode}; \code{required\_input\_artifact\_types}; \code{expected\_inputs} and \code{expected\_outputs}; \code{max\_turns}, optionally \code{\_by\_mode}; \code{loop\_hooks}, \code{loop\_hooks\_disable}, \code{hook\_config}; \code{summarizer} (trigger, what is kept verbatim, compression target); \code{callable\_nodes} (\code{[]} forbids \code{run\_node}, \code{["*"]} permits any); \code{post\_run\_flow} $\in$ \{\code{full}, \code{review\_curate}, \code{review\_only}, \code{none}\}; \code{shell\_probe\_only}; \code{deliverable\_writes}; and model overrides. An unknown \code{post\_run\_flow} value raises at load.

\paragraph{Terminal states.} A run ends in one of \code{completed}, \code{incomplete} (a declared output missing, or no committed change in scope), \code{blocked\_missing\_inputs} (a scientific precondition absent), \code{paused} (resumable in place), \code{cancelled} (external kill; nothing rolled back; the summary carries who requested it and at which turn), or \code{error}. The loop's terminal causes are: model finished; hook terminal (completed or failed); turn cap; external kill; degenerate repetition; protocol circuit break; progress circuit break; provider void; pause event.

The counts below are as of the repository on 2026-09-22. They are given for scale, not as a specification, and change as the code changes; \emph{Tools} is the length of the node's whitelist and \emph{Local skills} the number of skill files under the node's own directory (a further eight shared skills are declared by name).

\begin{table}[H]
\centering\footnotesize
\begin{tabular}{L{2.1cm}L{1.4cm}L{1.9cm}rrL{5.2cm}}
\toprule
Node & Role & Post-run flow & Tools & Local skills & Required outputs / callable nodes \\
\midrule
Analysis (\code{hypothesis}) & producing & full & 34 & 3 & preregistration, research plan, research state; calls \code{literature} \\
experiment & producing (high risk) & review, curate & 60 & 7 & experiment log, clean results, raw results; calls \code{data} \\
observation & producing & full & 20 & 0 & observation log with search protocol; calls \code{literature}, \code{data} \\
derivation & producing & review, curate & 28 & 0 & derivation log; calls \code{literature}, \code{data} \\
writing & producing & full & 28 & 7 & preflight plan, manuscript, validation report; calls \code{postprocess} \\
literature & service & none & 25 & 1 & per mode: survey report and index, or evidence package \\
data & service & none & 9 & 1 & dataset (custom loop) \\
postprocess & service & none & 16 & 23 & one figure record per visual request \\
\code{\_orchestrator} & architecture & -- & 32 & 0 & may call any node; sole holder of \code{run\_node}, decision packages, runtime control \\
\code{\_reviewer} & architecture & -- & 26 & 0 & review critique; loads the reviewed node's \code{review\_spec.md} \\
\code{\_curator} & architecture & -- & 36 & 0 & no artifacts; writes the knowledge base only \\
\bottomrule
\end{tabular}
\caption{The eleven node contracts as declared (counts as of 2026-09-22).}
\end{table}

\section{Record Shapes}
\label{app:records}

\paragraph{Artifact record} (one JSON file per artifact, path \path{<node>/artifacts/{type}__{slug}.json}): \code{type}, \code{name}, \code{content}, \code{metadata}, \code{created\_at}, \code{provenance} \{\code{kind} $\in$ produced $\mid$ imported $\mid$ forwarded, \code{by\_node\_type}, \code{by\_run\_id}, and for imports \code{source\_path}, \code{source\_sha256}, \code{source\_size\_bytes}\}, \code{version}, \code{content\_hash}, and when overwriting \code{prev\_content\_hash}; after an amendment, \code{amendment} \{\code{from\_version}, \code{reason}, \code{at}, \code{by\_node\_type}, \code{by\_run\_id}\}. Framework-owned metadata keys: \code{frozen}, \code{frozen\_at}, \code{freeze\_reason}, \code{review\_state}, \code{content\_assembled\_from}, \code{phantom\_claim\_ids}. Model-declared ledger keys: \code{measured\_metrics}, \code{closure\_discharges}, \code{run\_role}, \code{expected\_params}.

\paragraph{Freeze ledger rows} (\path{.frozen.jsonl}, append-only; every row after the first carries \code{prev\_row\_sha256}):
\begin{quote}\footnotesize
\code{\{"action":"freeze","artifact\_id","version","path","sha256","frozen\_at","by\_node","by\_run"\}}\\
\code{\{"action":"amend"|"migrate","artifact\_id","path","from\_version","to\_version","reason",}\\
\code{\ "snapshot\_path","snapshot\_sha256","diff":\{"changed\_metadata\_keys","content\_changed",}\\
\code{\ "content\_diff\_excerpt","old\_content\_hash","new\_content\_hash"\},"at","by\_node","by\_run"\}}
\end{quote}
A \code{freeze} row pins \code{path} to \code{sha256}; an \code{amend} or \code{migrate} row unpins \code{path} and pins \code{snapshot\_path} to \code{snapshot\_sha256}. The diff excerpt is capped at 4000 bytes and excludes the framework-owned freeze keys.

\paragraph{Memory entries} (\path{MEMORY.md}, sections delimited by \code{<!-- section:\{name\} -->} for \code{goal}, \code{law}, \code{narrative}, \code{manual\_pitfall}, \code{manual\_method}): a bullet followed by a comment line. A law: \code{<!-- @ derived=N | at=DATE | src=... -->}. A handbook entry: \code{<!-- @ tools=a,b | nodes=x | run=RUN | commit=SHA | seen=N | last=DATE | defect=0|1 -->}. The constitution sections are capped at 2048 bytes each; near-duplicate detection uses word-shingle Jaccard $\ge 0.75$ or an identical 8-token prefix; a recurrence count of 5 flags a defect.

\paragraph{Knowledge card} (organization tier): \code{domain}, \code{statement}, \code{applicability}, \code{why}, \code{practice}, \code{confidence\_basis}, \code{evidence}; for dead ends also \code{trigger}, \code{cost\_when\_hit}; plus \code{promoted\_from} \{\code{project\_id}, \code{source\_id}, \code{approved\_by}, \code{at}\}, \code{org\_kind}, \code{replication\_count}. Identifiers are \code{\{entity\}\_\{sha12(signature|scope)\}}.

\paragraph{Cost ledger row} (\path{.harness/llm_cost.jsonl}, one per model call): \code{ts}, \code{run\_id}, \code{node}, \code{provider} (host and port), \code{model}, \code{turn}, \code{prompt\_tokens}, \code{completion\_tokens}, \code{total\_tokens}, \code{cache\_read}, \code{cache\_write}, \code{cache\_hit\_ratio}, \code{cost\_usd}, \code{price\_known}. Unreported cache fields are \code{null}, never 0; \code{cost\_usd} is \code{null} when \code{price\_known} is false.

\section{Delivery Channels and Budgets}
\label{app:channels}

\begin{table}[H]
\centering\small
\begin{tabular}{L{3.4cm}L{4.2cm}L{5.8cm}}
\toprule
Channel & Trigger & Budget \\
\midrule
constitution & every turn, all nodes & source capped at 2048 bytes \\
situation & each run start & 3072 bytes \\
onboarding slice & turn one, all nodes & 4096 bytes, matched on \code{applies\_to} \\
first-use tool brief & first call of a given tool & at most 2 entries, 512 bytes \\
knowledge-base opening injection & project opening & constant: 10 conclusions, 5 dead ends \\
\bottomrule
\end{tabular}
\caption{Mechanical delivery channels. None is gated on a query; every budget is a constant independent of the size of the store behind it.}
\end{table}

\section{Reproduction: Where the Numbers Come From}
\label{app:repro}

All figures in this paper are computed from records the platform writes as a side effect of running; none were produced for the paper.

\begin{itemize}
\item \textbf{Case C1.} Cost ledger: \path{<data-root>/projects/46da60b0-.../.harness/llm_cost.jsonl} (2,021 rows). Freeze record: \path{writing/artifacts/.frozen.jsonl} (manuscript version 11, \code{frozen\_at} 2026-09-01T14:46:01Z). Review critiques: \path{reviews/artifacts/review_critique__*.json} (20 files with \code{metadata.verdict} and \code{metadata.confidence}). Research state: \path{hypothesis/artifacts/research_state__research_state.json} (version 9). Figure~\ref{fig:c1} is a direct rendering of the ledger and the critique timestamps.
\item \textbf{Case C2.} \path{deliverables/e2e_v26_ising_tc_20260813/}: frozen manuscript PDF, \path{raw_results.json} (800 records), \path{clean_results__*.json}, \path{experiment_log__*.json} (frozen, with verdict reasoning), \path{research_state__v1..v5.json}, simulation source, and the 8.85-hour progress log. The theory-experiment chain: \path{benchmarks/derivation/T_LINE_RESULT_20260823.md} and \path{scripts/verify_prediction_precedes_experiment.py}. The compounding replay: \path{scripts/replay_ising_closure.py} and \path{docs/KB_CLOSURE_PROOF_20260821.md}.
\item \textbf{Verdict-layer audit} (Section~\ref{sec:noverdict}): \path{docs/PROPOSAL_RETIRE_QC_AS_A_VERDICT_LAYER_20260810.md}.
\item \textbf{Cost} (Section~\ref{sec:cost}): the Ising project's \path{llm_cost.jsonl}, cut into contiguous node segments; the Monte Carlo interval is the gap with no model calls.
\item \textbf{Scale counts} (as of 2026-09-22). Tools: the union of the whitelists declared in the eleven contracts (140). Skills: files named \code{SKILL.md} under \path{nodes/} (42) and \path{shared/} (8).
\end{itemize}